%% file: main.tex
\documentclass[sigplan,nonacm]{acmart}

{}
\newcommand{\name}{TranSAGE}
\newcommand{\rev}[1]{#1}

\usepackage{amsmath,amsfonts}
\usepackage{graphicx}
\usepackage{textcomp}
\usepackage{xcolor}
\usepackage{multirow}
\usepackage{listings}
\usepackage{booktabs}
\usepackage{tabularx}
\usepackage{multicol}
\usepackage{enumitem}
\usepackage{tikz}
\usepackage{subcaption}
\usepackage{rotating}
\usepackage{pifont}
\usepackage{adjustbox}
\usepackage{svg}
\usepackage{makecell}
\usepackage{threeparttable}
\usepackage{seqsplit}
\usepackage{algorithm}
\usepackage{algpseudocode}
\usepackage{balance}

\author{Chi Zhang}
\affiliation{%
  \institution{Shanghai Jiao Tong University}
  \city{Shanghai}
  \country{China}
}
\email{zhang-chi@sjtu.edu.cn}

\author{Yi Sun}
\affiliation{%
  \institution{School of Microelectronics, South China University of Technology}
  \city{Guangzhou}
  \country{China}
}
\email{misunyi@mail.scut.edu.cn}

\author{Weichuang Zhang}
\affiliation{%
  \institution{Shanghai Jiao Tong University}
  \city{Shanghai}
  \country{China}
}
\email{1064080006@sjtu.edu.cn}

\author{Quan Chen}
\affiliation{%
  \institution{Shanghai Jiao Tong University}
  \city{Shanghai}
  \country{China}
}
\email{chen-quan@cs.sjtu.edu.cn}

\author{Jieru Zhao}
\authornote{Corresponding author}
\affiliation{%
  \institution{Shanghai Jiao Tong University}
  \city{Shanghai}
  \country{China}
}
\email{zhao-jieru@sjtu.edu.cn}

\begin{document}
\title[\name{}]{\name{}: Automating \underline{S}patial \underline{A}ccelerator \underline{Ge}neration for Large-Scale \underline{Tran}sformer Models}

\begin{abstract}

Large-scale Transformers have achieved remarkable success across NLP, computer vision, and robotics, but their growing computational and memory demands challenge efficient inference.
Spatial accelerators offer advantages in on-chip bandwidth and energy efficiency, yet designing them for large models remains costly and time-consuming. Existing accelerator generators typically focus on single-operation support, overlooking the benefits of multi-operation pipelining, which limits throughput and energy efficiency.

We propose TranSAGE, a compiler-based framework that jointly optimizes scheduling for large-scale transformers under multi-operation pipelining and generates corresponding accelerator implementations. 
Our approach integrates a DSE engine with multi-operation dataflow optimization and a dynamic programming-based graph partitioning algorithm to efficiently identify pipeline-friendly operation clusters and their optimal dataflow parameters. 
We further design spatial architecture primitives and lowering passes to automatically generate efficient RTL implementations.

Evaluations on four representative LLMs demonstrate that TranSAGE achieves up to 2.41$\times$ higher throughput and 13.04$\times$ higher energy efficiency over commercial GPU/NPU baselines. Compared with state-of-the-art accelerator generators, TranSAGE achieves up to 11.33$\times$ higher throughput and 6.07$\times$ higher energy efficiency. Our code is available at \url{https://github.com/sjtu-zhao-lab/TranSAGE.git}.

\end{abstract}


\maketitle 

\input{section/intro}

\input{section/background}

\input{section/overview}

\input{section/frontend}
\input{section/backend}
\input{section/evaluation}

\input{section/discussion}

\input{section/conclusion}

\begin{acks}
This work was supported by the National Natural Science Foundation of China under Grant 62622214 and Grant 62472273. Codex were used for language and code assistance.
\end{acks}

\input{section/appendix}

\input{section/ae}

\clearpage

\bibliographystyle{ACM-Reference-Format}
\balance
\bibliography{acmart}

\end{document}

%% file: section/intro.tex
\section{Introduction}
\label{sec:introduction}

Large-scale transformers, including large language models (LLMs), Diffusion Transformers (DiTs), vision-language models (VLMs), and vision-language-action (VLA) models, have demonstrated remarkable capabilities across a broad spectrum of applications, such as conversational agents \cite{ATTENTION,BERT,LLM_EMERGENCE,SCALING_LAW}, multimedia content generation \cite{DDPM, DIT, VIDEO_DIT, LATENT_DIT}, multimodal understanding \cite{CLIP, BLIP}, and robotics \cite{RT1, PALME}. 
\rev{However, the continued growth in model scale increases the computational and memory demands of these models, making low-latency and energy-efficient inference increasingly challenging.}
These challenges are further amplified by inference-time scaling techniques, such as Chain-of-Thought (CoT) reasoning \cite{COT} and agent-based methods \cite{REACT, AUTOGPT}.

The challenges have driven the need for architectural innovations. Although GPUs remain the most widely adopted platform for deploying such models, they suffer from limited energy efficiency and suboptimal throughput.
Recent research has increasingly focused on spatial accelerators to improve the efficiency of large-scale transformer inference \cite{SOFA, SHIDIANNAO, TPU, ASCEND, EYERISS, GROQ}. 
Unlike conventional architectures that heavily rely on deep memory hierarchies and temporal locality—often leading to bandwidth bottlenecks—spatial accelerators emphasize distributed memory and localized point-to-point communication. This design significantly enhances on-chip bandwidth and aligns well with the computational characteristics of transformer models. In particular, tensor operations in transformers exhibit high spatial data reuse with minimal need for global synchronization, making them especially well-suited for spatial execution paradigms.


\begin{table*}[]
  \centering
  \caption{Comparison with representative spatial accelerator automation frameworks.}
  \label{tab:generator_comp}%
  \resizebox{\textwidth}{!}{
  \begin{threeparttable}
  
    \begin{tabular}{ccclcccccc}
    \toprule
    \multirow{2}[4]{*}{Work} & \multicolumn{2}{c}{\textbf{Productivity}} &       & \multicolumn{2}{c}{\textbf{Dataflow Support}} &       & \multicolumn{3}{c}{\textbf{Architecture Support}} \\
\cmidrule{2-3}\cmidrule{5-6}\cmidrule{8-10}          
          & Input & \makecell{Auto-Tuning \\ Support} &       
          & \makecell{Adaptive \\ Dataflow} & \makecell{Inter-Op \\ Dataflow} &       
          & PU &  \makecell{Intra-PU \\ Topology} &  \makecell{Inter-PU \\ Topology} \\
    \midrule
    AutoSA \cite{AUTOSA}
          & C Code 
          & \ding{51}     
          &       
          & \ding{55}   
          & Sequential 
          &       
          & SA \footnote[2]{} Only
          & Fixed & / \\
    TensorLib \cite{TENSORLIB}
          & STT \footnote{a}
          & \ding{55}       
          &       
          & \ding{55}  
          & Sequential 
          &       
          & SA \footnote[2]{} Only
          & Fixed & / \\
    Gemmini \cite{GEMMINI}
          & C Code
          & \ding{55}       
          &       
          & \ding{55}  
          & Limited Fusion 
          &       
          & SA \footnote[2]{} + PPU \footnote[3]{}
          & Fixed & Fixed \\
    LEGO  \cite{LEGO}
          & Relation Matrix 
          & \ding{55}       
          &       
          & \ding{51}  
          & Limited Fusion 
          &       
          & SA \footnote[2]{} + PPU \footnote[3]{}
          & Reconfigurable  & Fixed \\
    TranSAGE (Ours) 
          & PyTorch Code 
          & \ding{51}       
          &       
          & \ding{51}  
          & Tile-based Pipeline
          &       
          & Reconf. PU
          & Reconfigurable  & Flexible \\
    \bottomrule
    \end{tabular}%

\begin{tablenotes}[para]
\item[1] Spatial-Temporal Transformation
\item[2] Systolic Array 
\item[3] Post Processing Unit
\end{tablenotes}

\end{threeparttable}
}
\end{table*}%

However, maximizing the performance of large-scale transformer models on spatial accelerators requires substantial expertise in both hardware and software and is highly time-consuming, thereby hindering their broader adoption. 
\rev{Numerous mapping and compiler frameworks have been proposed to identify optimal mappings of large-scale neural networks onto spatial accelerators \cite{SET, TILEFLOW, TIMELOOP, MAESTRO, TENET, CRANE, DSAGEN, SPATIAL, SPARSEABM, CODO, STREAMTENSOR, STREAMHLS, APEX}.}
Although these frameworks report promising results, they primarily target predefined architectures with fixed or limited configurability, and thus lack the ability to jointly explore optimizations across both mapping strategies and architectural designs.


In contrast, a growing body of automated frameworks has emerged to generate optimized architectures and scheduling strategies for target workloads. We summarize these approaches in Table~\ref{tab:generator_comp}. AutoSA \cite{AUTOSA} proposes a polyhedral-based framework that automatically transforms C code into optimized HLS implementations for constructing systolic arrays. TensorLib \cite{TENSORLIB} introduces a novel mathematical representation of spatial-temporal transformations (STT), and generates input, output, and interconnect components based on analysis of the STT matrix using Chisel \cite{CHISEL} templates. However, they are primarily designed for systolic arrays, which can limit their efficiency on diverse operations in large-scale transformer models.

Gemmini \cite{GEMMINI} proposes a system-integrated, template-based generator that employs a systolic array as the core matrix processing unit, augmented with post-processing units to support a broader range of operations. LEGO \cite{LEGO} further extends this architecture by introducing adaptive dataflows on systolic arrays to mitigate the low utilization caused by rigid dataflow mappings. Although they target compound workloads, the generated accelerators may still incur high memory access overhead and achieve limited throughput. This limitation primarily arises from their support for only fixed-pattern fusion (e.g., GEMM combined with simple element-wise operations), which restricts their ability to fully exploit cross-operator optimization opportunities—such as operation fusion and pipelining—that have been widely demonstrated to be effective \cite{LAYERFUSE, PIPEFUSER, MULTIFUSE, STREAM, TILEFLOW, SET}.

\textit{\textbf{Main idea.}} The root cause of the inefficiency in the aforementioned generators is that they treat individual operations as the basic execution unit and generate architecture for supporting them. In contrast, our key idea is to elevate clusters to first-class citizens rather than individual operations. Specifically, we \rev{fuse} operations that are well-suited for parallel execution into clusters, which are then generated as hardware pipelines. The execution of the entire model is thus decomposed into the sequential execution of multiple clusters. Meanwhile, all clusters share and reuse the same on-chip compute and memory resources.


\textbf{\textit{Challenge.}} However, such a design is challenging due to two aspects.
\underline{(1)} \textit{Difficulty in identifying optimal dataflow:} For large-scale transformer models, which typically consist of dozens of operations, determining which operations should be \rev{fused} into a pipeline cluster is non-trivial. Furthermore, even after \rev{fusion}, it remains challenging to assign dataflow parameters (e.g., tiling sizes, unrolling factors, and execution ordering) for each operation within a complex pipeline, making it difficult to identify the optimal dataflow. We elaborate on how these factors expand the design space and introduce potential pitfalls in Section~\ref{sec:motivation}.
\underline{(2)} \textit{Difficulty in generating flexible architectures:} Even after determining the optimal dataflow for compound operations, generating corresponding hardware architectures remains challenging. It is impractical to implement architectures for every pipeline configuration, whether manually or using template-based methods. Moreover, all clusters must be mapped onto constrained hardware resources while minimizing overhead, further complicating the design.

\textbf{\textit{Our solution.}} To address these challenges, we propose \name{}, an MLIR-based compiler framework that automatically extracts optimal operation clusters with corresponding dataflow parameters for large-scale transformer models, and generates the associated RTL implementations.

We introduce a design space exploration (DSE) engine that 
formulates multi-operation pipeline scheduling as two subproblems: graph partitioning and dataflow optimization. For graph partitioning, \name{} employs a dynamic programming–based method to determine the optimal \rev{operation fusion} under a given topological order (Sec. \ref{sec:operation_clustering}). For each cluster, \name{} searches for the optimal dataflow parameters of its operators to form a pipelined cluster, using a decomposition-based approach (Sec. \ref{sec:intracluster_dataflow}). 


Furthermore, we propose a set of spatial architecture primitives (SAPs) that abstract arithmetic, data storage, and routing components commonly found in spatial accelerators. These primitives serve as a bridge between high-level representations (e.g., \textit{MLIR linalg}) and RTL implementations. Based on dependency and data-sharing analysis, high-level linalg operations are lowered to SAPs and subsequently mapped to corresponding RTL implementations (Sec. \ref{sec:back_end}).


Our main contributions are summarized as follows:
\begin{itemize}
    \item We propose \name{}, an end-to-end framework that automatically generates RTL implementations of spatial accelerators from PyTorch programs. 
    \item We integrate a DSE engine that automatically extracts optimal \rev{operation fusion} and dataflow parameters, enabling efficient mapping to hardware pipelines for maximum performance.
    \item We propose a set of abstractions for spatial accelerator hardware components, along with corresponding lowering passes, which serve as a bridge between high-level IR and RTL implementations.
\end{itemize}


\rev{Evaluation shows that \name{} delivers up to 2.42$\times$ higher throughput and 13.04$\times$ greater energy efficiency than commercial GPU/NPU baselines.} Against state-of-the-art accelerator generators, \name{} further improves throughput by up to 11.33$\times$ and energy efficiency by up to 6.07$\times$.

%% file: section/background.tex
\section{Background and Motivation}

\subsection{Spatial Accelerators}

\begin{figure}[]
    \centering
    \includegraphics[width=\columnwidth]{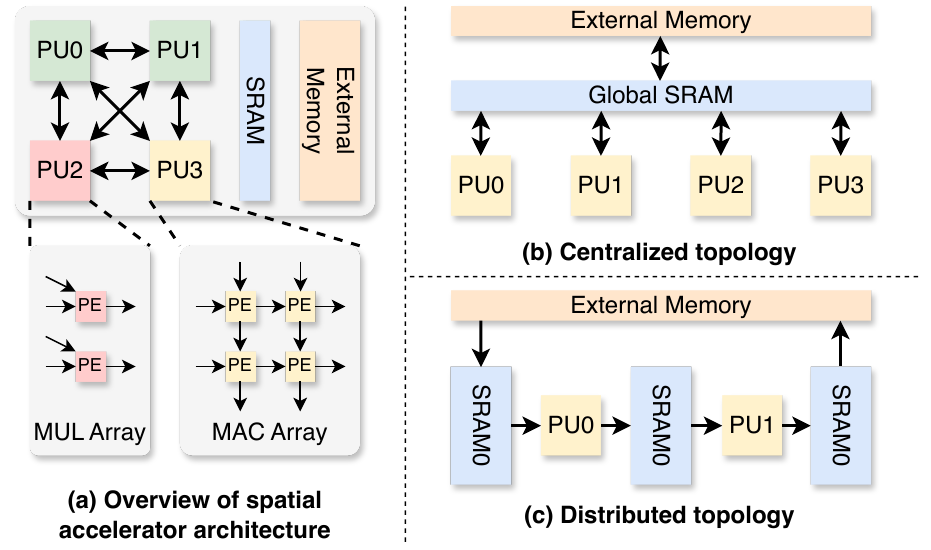} 
    \caption{Spatial accelerator architecture.}
    \label{fig:arch_spatial_accelerator}
\end{figure}


\textbf{\textit{Spatial accelerators}} represent a broad class of architectures characterized by two key features: \textit{(1) distributed compute and memory resources} and \textit{(2) asynchronous communication patterns}. These characteristics manifest across multiple levels of the system hierarchy. At the processing element (PE) level, designs such as Eyeriss~\cite{EYERISS} and TPU~\cite{TPU} organize PEs with local register files and localized interconnects (e.g., systolic arrays), enabling data reuse across neighboring PEs while reducing the overhead associated with global synchronization. At higher levels, such as the core or package level, systems like Simba~\cite{SIMBA} and Cerebras~\cite{CEREBRAS} preserve these principles while scaling out through network-on-chip (NoC) or network-on-package (NoP) interconnects.



\textbf{\textit{\name{} Scope.}} The architectural scope considered in \name{} is illustrated in Fig.~\ref{fig:arch_spatial_accelerator} (a). Within this scope, we focus on a single compute core equipped with on-chip compute resources and a hierarchical memory system. As discussed earlier, this design can be further scaled through inter-core or inter-package interconnects. The design space of spatial accelerators within this scope consists of two primary dimensions. \textit{(1) Organization of PEs.} For each type of compute resources (e.g., multipliers, adders, and nonlinear arithmetic units), we must determine the size of the PE array and how PEs are connected to local registers and to each other. \textit{(2) Interconnect of PUs.} Furthermore, we need to define the topology of processing units (PUs). One option is a centralized topology, as shown in Fig.~\ref{fig:arch_spatial_accelerator}(b), where all PUs are connected to a global SRAM, enabling flexible data exchange. Alternatively, specialized PU topologies can be employed, as illustrated in Fig.~\ref{fig:arch_spatial_accelerator}(c), where connectivity between PUs is restricted to reduce communication overhead.

\subsection{Dataflow and Representation}
\label{sec:dataflow_background}
Dataflow refers to the scheduling strategy that describes how tensor operations are temporally and spatially mapped onto a spatial accelerator. It typically specifies the size and order of data movement between components, both within the same hierarchy level and across different hierarchy levels. In addition, it characterizes how computations are executed, whether in parallel or sequentially.

\begin{figure}[]
    \centering
    \includegraphics[width=\columnwidth]{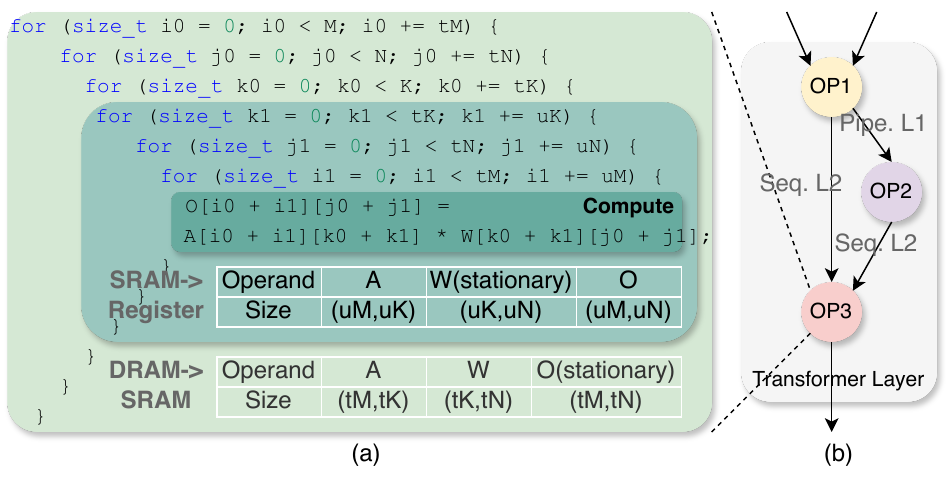} 
    \caption{Representation for (a) intra-operation dataflows and (b) inter-operation dataflows. Seq. and Pipe. denote sequential and pipelined execution, and L1 and L2 denote data transfers through SRAM and DRAM, respectively.
    }
    \label{fig:dataflow_background}
\end{figure}


\textbf{\textit{Intra-operation dataflow and representation.}} In the early research stage, tensor operations such as matrix multiplication and convolution are executed sequentially on spatial accelerators. In this setting, it is sufficient to represent the dataflow of each operation independently, which we refer to as intra-operation dataflow. The most commonly used method for representing intra-operation dataflow is a nested loop with annotations. Fig.~\ref{fig:dataflow_background}(a) shows an example of the dataflow for a matrix multiplication operation on a spatial accelerator with a two-level memory hierarchy.

As illustrated, the six-level nested loop is derived from the original three-level nested loops over the $M$, $N$, and $K$ dimensions. Specifically, for data movement across memory hierarchy, the representation explicitly captures the data size through tiling along each dimension. For example, the tiling at the first three levels represents data transfer between DRAM and SRAM with a size of $(tM, tN, tK)$, while the rest levels represent data transfer between SRAM and registers with a size of $(uM, uN, uK)$. We refer to these as the \textbf{tiling vector} and\textbf{ unroll vector}, respectively.

Furthermore, within each hierarchy, the ordering of the three dimensions is a permutation of the original $(M, N, K)$ order, which determines the access pattern of each tensor. For instance, an order of $(M, N, K)$ at the first three levels implies that $A(m,k)$ and $W(k,n)$ are accessed in a strided manner, while $O(m,n)$ is accessed contiguously, indicating that $O$ is stationary in SRAM. Similarly, at the rest levels, $W(k,n)$ exhibits a stationary pattern in the register file. We encode these permutations as index vectors relative to the original order, referred to as the \textbf{outer-order vector} and \textbf{inner-order vector} for the first three and rest levels, respectively.


\textbf{\textit{Inter-operation dataflow and representation.}}
\rev{As techniques such as layer fusion and pipeline execution have demonstrated advantages in reducing external memory access and improving PE utilization \cite{DNNFUSION, FUSEMAX, TRANSFUSION, TILUS, OPFUSEDAC25}, increasing attention has been devoted to representing dataflows that span multiple operations, i.e., inter-operation dataflow.}
For example, methods in \cite{SET, TILEFLOW} adopt tree-based representations with annotations to describe how resources are temporally and spatially allocated across operations. In contrast, the table-based method in \cite{CRANE}
provides a more formal mathematical abstraction.
Despite differences in representation, the key objective of inter-operation dataflow modeling is to capture how data is transferred between operations (e.g., through DRAM, SRAM, or registers) and how operations are scheduled (e.g., sequentially, in parallel, or in a pipeline). Therefore, a directed acyclic graph (DAG) with annotated edges is often a natural and effective choice for representing inter-operation dataflow.
As illustrated in Fig.~\ref{fig:dataflow_background}, the DAG shows that OP1 and OP2 are pipelined through SRAM, while both execute sequentially with OP3 via DRAM.

\subsection{Issues and Motivation}
\label{sec:motivation}

As discussed in Sec.~\ref{sec:introduction}, 
the common limitation of existing spatial accelerator generators is the lack of support for efficient multi-operation dataflows.
Specifically, they either target a single operation on a single systolic array or support fixed fusion patterns such as GEMM followed by activation. Such designs are insufficient for large-scale transformer models, which consist of complex compositions of diverse operations with varying topologies. As a result, these generators often execute operations sequentially, leading to underutilization of PEs and incurring significant memory access overhead.

To address this limitation, our key idea is to group multiple operations and execute them in a pipelined manner, thereby maximizing throughput while minimizing external memory accesses. Instead of generating hardware for individual operations, we propose synthesizing architectures tailored to such operation clusters. However, this approach introduces several issues in designing efficient spatial accelerators.

\begin{figure}[]
    \centering
    \includegraphics[width=\columnwidth]{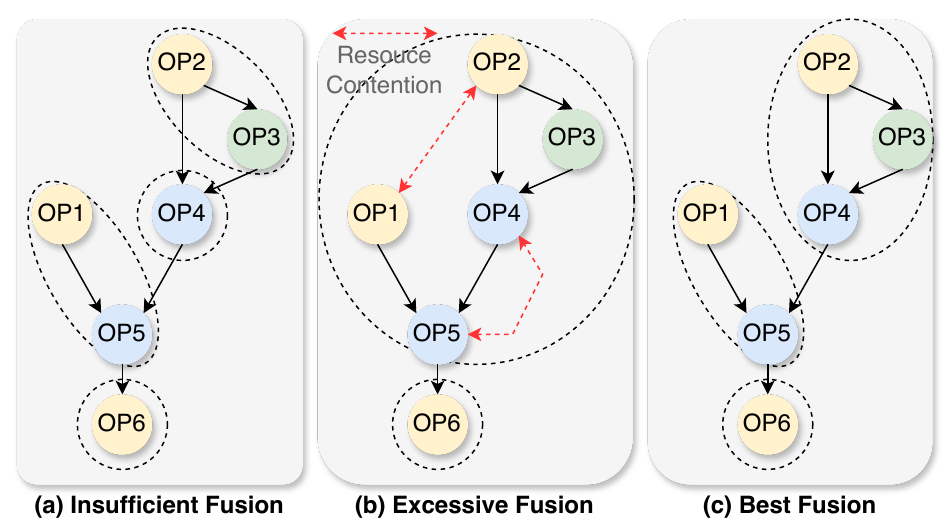} 
    \caption{Example of different \rev{fusion} results.}
    \label{fig:cluster_motivation}
\end{figure}

\textit{\textbf{Pitfall 1: Suboptimal \rev{operation fusion}.}}
Given the large computation graph of transformers, determining how to \rev{fuse} operations for optimal performance is highly challenging. Both insufficient \rev{fusion} and excessive \rev{fusion} can lead to suboptimal results. Fig.~\ref{fig:cluster_motivation} illustrates three \rev{fusion} configurations explored by our DSE engine (to be detailed in Sec.~\ref{sec:front_end}) for the FFN layer in the prefill stage of LLaMA3-8B. 
Configuration (a) represents insufficient \rev{fusion}: although OP2, OP3, and OP4 utilize different compute resources, they are divided into separate clusters. 
In contrast, (b) illustrates excessive \rev{fusion}, where OP1–OP5 are \rev{fused} into a single cluster. Under pipelined execution, operations within the same cluster compete for compute and storage resources (e.g., SRAM), leading to what we term resource contention. This results in PE underutilization for two reasons: (1) imperfect PE allocation cannot fully utilize available resources. For example, assuming a total of 64 PEs, no valid allocation can fully utilize all PEs for two pipelined operations with a FLOP ratio of 2:1; and (2) reduced data reuse, because the limited per-operation SRAM capacity prevents each operation from storing sufficient inputs and partial sums, potentially pushing the computation into a memory-bound regime.


Interestingly, the optimal \rev{fusion} strategy is workload-dependent. When the workload shifts from the prefill stage to the decoding stage, the same computation graph may favor configuration (b),
because \rev{fusing} two memory-bound operations improves overall resource utilization.

\textit{\textbf{Pitfall 2: Large and complex dataflow design space.}}
Even with an ideal operator \rev{fusion}, identifying the optimal dataflow parameters for each cluster remains challenging due to the enormous design space. Consider a cluster containing $N$ operations, denoted as $OP_i$, where each operation is represented by an $M_i$-level nested loop with bounds $(b_{i0}, b_{i1}, \ldots, b_{iM_i-1})$. As introduced in Section~\ref{sec:dataflow_background}, the dataflow design for each $OP_i$ requires determining: (1) a tiling vector $(t_{i0}, t_{i1}, \ldots, t_{iM_i-1})$, with a search space of $\prod_{j=0}^{M_i-1} b_{ij}$; (2) an unrolling vector $(u_{i0}, u_{i1}, \ldots, u_{iM_i-1})$, which, for each tiling configuration, has a search space of $\prod_{j=0}^{M_i-1} t_{ij}$; (3) an outer loop ordering, with a space of $M_i!$; and (4) an inner loop ordering, also with a space of $M_i!$. 

When combined across all operations in the cluster, the total design space grows exponentially.
In practice, for a cluster of six operations (our default maximum cluster size in experiments) in a sequence-length-1024 attention layer, the overall design space can reach up to $10^{21}$ possible configurations.


\textit{\textbf{Pitfall 3: Difficulty in generating hardware for diverse operation clusters.}}
After scheduling, another major challenge lies in generating RTL code for the extracted clusters. As discussed earlier, the resulting clusters may exhibit diverse topologies, along with varying SRAM and PE resource allocations. Such diversity significantly complicates RTL development. Moreover, conventional approaches such as hardware templating are difficult to apply due to the lack of flexibility required to accommodate these heterogeneous designs. Furthermore, all clusters must ultimately be mapped onto a unified architecture with shared resources, introducing additional constraints and design complexity.



%% file: section/overview.tex
\section{Framework Overview}
\label{sec:framework_overview}

To address all the challenges, we propose \name{}, an automated framework that takes PyTorch code as input, automatically performs multi-kernel pipeline scheduling to extract clusters with optimized dataflow, and generates RTL code for hardware implementation. 
The overview is shown in Fig.~\ref{fig:framework_overview}. 
The framework follows a two-stage workflow.




\textbf{Generation stage}. 
\name{} first takes the PyTorch \cite{TORCH} code of the input model and uses Torch-MLIR to generate the corresponding MLIR representations.
The frontend begins with Linalg canonicalization and optimization, transforming the input IR into a representation composed solely of \texttt{linalg.generic} operations and auxiliary tensor operations. Based on it, \name{} fuses memory-oriented generic operations to eliminate redundant memory accesses.
\name{} then performs multi-operation pipeline scheduling with a DSE engine to identify optimal operation fusion and corresponding dataflow parameters (\textit{Challenge 1}). 
Afterwards, \name{} performs backend lowering to generate binary and RTL code from the Linalg IR of these fused clusters (\textit{Challenge 2}).
We introduce a set of spatial architecture primitives as a bridge between Linalg and RTL implementation. 
Based on these primitives, we construct a primitive graph that maps all supported clusters onto shared hardware and lowers it to binary and RTL code. 
\rev{Notably, different shapes may favor different schedules. Therefore, \name{} independently runs the frontend for representative shapes, incorporating all distinct clusters into the mapping stage.
Details are described in Sec. \ref{sec:back_end}.}

\rev{\textbf{Inference stage.} To serve models on the generated accelerator, \name{} uses a host runtime with a dispatch table that maps input shapes to binary code. 
For each request, the runtime selects the binary for the corresponding shape range. It then fills in the runtime loop bounds and memory addresses, supporting any shape within the range without architectural modification.
For example, as shown in Figure \ref{fig:framework_overview}, a request with a sequence length of 493 uses \textit{Binary Code1}, which corresponds to the sequence length range from 256 to 512. The runtime then configures the accelerator accordingly.}






%% file: section/frontend.tex
\section{Multi-Operation Pipeline Scheduling}
\label{sec:front_end}

\textbf{\textit{Problem Formulation.}} 
The canonicalized Linalg IR consists of \texttt{linalg.generic} operations connected by tensor operations. 
During multi-operation pipeline scheduling, we partition these generic operations into clusters, each representing a candidate pipelined grouping.
For each cluster, we characterize the pipelined dataflow using parameters including the tiling vector, unrolling vector, and outer/inner loop ordering (Sec.~\ref{sec:dataflow_background}), which define the execution behavior of each cluster.
Therefore, the scheduling problem becomes a joint optimization over \rev{fusion} schemes and dataflow parameters, aiming to find a configuration that optimizes the target metric, namely, \textit{energy delay product (EDP)}. 

\begin{figure}[]
    \centering
    \includegraphics[width=\columnwidth]{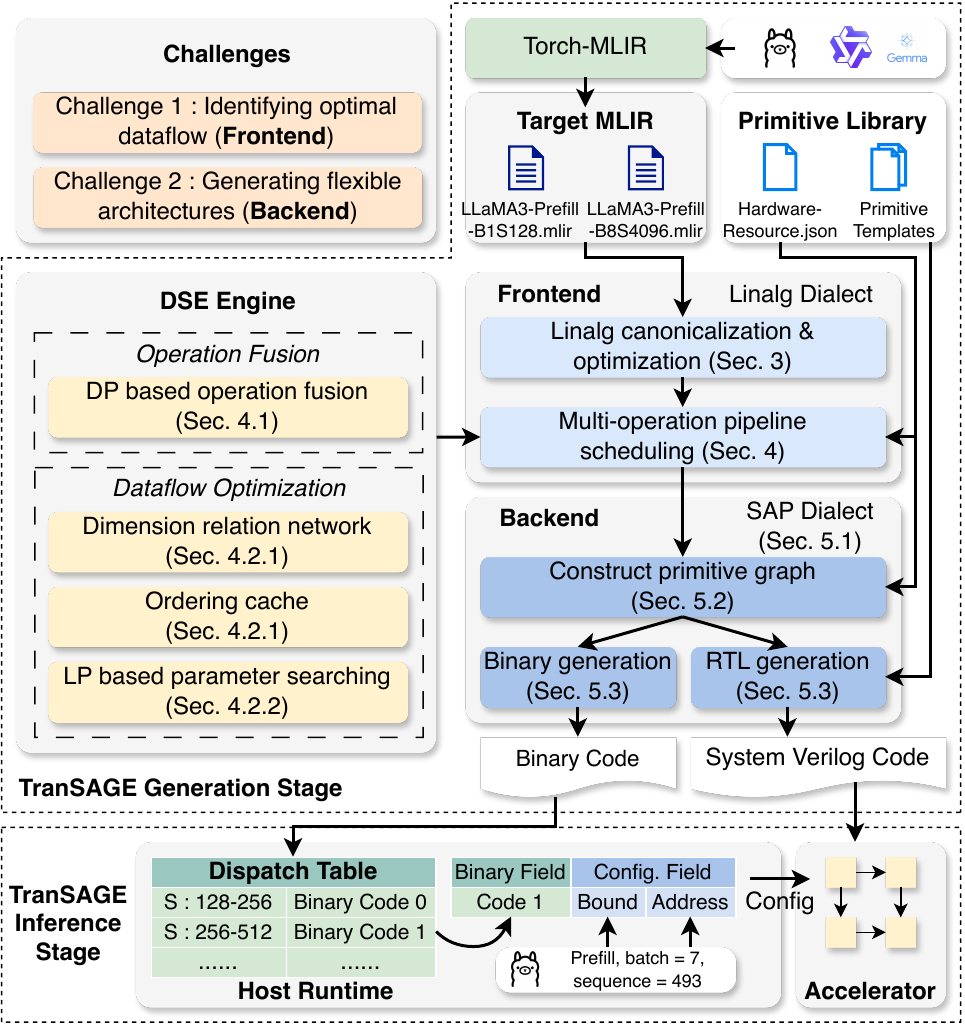} 

    \caption{\rev{Framework overview. }}
    \label{fig:framework_overview}
\end{figure}

\subsection{Operation \rev{Fusion}}
\label{sec:operation_clustering}




We formulate operation \rev{fusion} as a graph partitioning problem, where the objective is to partition the computation graph into several clusters. 
This problem has a natural recursive structure: given a topologically ordered sequence of operations, once the optimal final cluster is determined, it can be removed, reducing the original problem to a smaller subproblem over the remaining operations.


Therefore, we adopt a dynamic programming (DP)-based method for graph partitioning. First, all operations are sorted in topological order. For a total of \(N\) operations, we define an array \(\texttt{dpStatus}[0 \ldots N]\), 
where \(\texttt{dpStatus}[i]\) records the optimal EDP
for partitioning the subgraph consisting of operations \(0\) to \(i\). 
We also maintain an array \(\texttt{cutIndex}[0 \ldots N]\), where \(\texttt{cutIndex}[i]\) records the corresponding partition point. 
Our graph partitioning algorithm is presented in Algorithm~\ref{alg:clustering}. Starting from the first node, for each node \(i\), the algorithm enumerates all possible cut points \(j \leq i\) to identify a partition whose operations form a fully connected cluster. This is because fusing operations without data dependencies provides no theoretical benefit. Among valid partitions, 
the algorithm selects the one that minimizes the overall EDP. The corresponding optimal EDP is stored in \(\texttt{dpStatus}[i]\), while the selected cut point is recorded in \(\texttt{cutIndex}[i]\). 
After processing all nodes, we reconstruct the final partitioning by backtracking. Starting from \(i=N\), where \(i\) denotes the current cut index, we iteratively set \(i=\texttt{cutIndex}[i]\) to obtain the preceding cut index until reaching the beginning of the graph.
During backtracking, the optimal dataflow parameters for each cluster are determined (Sec. \ref{sec:intracluster_dataflow})
and attached to the corresponding operations as attributes, enabling subsequent stages to leverage them for RTL generation.

\begin{algorithm}[t]
\caption{DP-based Operation \rev{Fusion}}
\begin{algorithmic}[1]

\Require Operation list $[0 \dots N-1]$, max cluster size $K$
\Ensure Cutting index for optimal fusion and metrics

\State Initialize $dpStatus[0] \gets \infty$
\State $cutIndex[0] \gets 0$

\For{$i = 1$ to $N$}
    \State $bestEDP \gets \infty,\quad bestIndex \gets 1$


    \For{$j = 1$ to $i$}
        \State $cluster \gets$ operations $[j, i]$

        \If{connectivity($cluster$) $\land$ $i-j + 1 <= K$}
            \State $EDP \gets$ \textsc{Solve}($cluster$)
            \State $newEDP \gets \textsc{Merge}(dpStatus[j-1], EDP)$
    
            \If{$newEDP < bestEDP$}
                \State $bestEDP \gets newEDP$
                \State $bestIndex \gets j$
            \EndIf
        \EndIf
    \EndFor

    \State $dpStatus[i] \gets bestEDP$
    \State $cutIndex[i] \gets bestIndex$

\EndFor

\end{algorithmic}
\label{alg:clustering}
\end{algorithm}

\subsection{Intra-Cluster Dataflow Optimization}
\label{sec:intracluster_dataflow}



The objective at this stage is to determine the tiling vector, the unroll vector, the outer-order vector and the inner-order vector for each operation (Sec. \ref{sec:dataflow_background}) within a cluster to maximize the performance. 
As discussed in Sec.~\ref{sec:motivation}, the design space is extremely large, making brute-force search infeasible. 
However, we observe that unroll vectors and inner-order vectors of different operators are relatively independent (corresponding to the higher memory hierarchy \emph{L1} between SRAM and registers), whereas tiling vectors and outer-order vectors are tightly coupled (corresponding to the lower memory hierarchy \emph{L2} between DRAM and SRAM). Therefore, we propose a decomposed optimization approach. We first enumerate all possible combinations of tiling vectors and outer-order vectors, which we refer to as \emph{L2 parameters}. For each L2 configuration, we determine the optimal unroll factors and inner orders—denoted as \emph{L1 parameters}—using a well-formulated and efficient solving method. 

\subsubsection{Search Space Pruning for L2 Parameters.}
Despite the decomposition, the search space of \emph{L2 parameters} remains prohibitively large. To this end, we prune the search space through: a \textit{dimension-relation network representation} and an \textit{equivalence-based ordering cache}.

\begin{figure*}[]
    \centering
    \includegraphics[width=\textwidth]{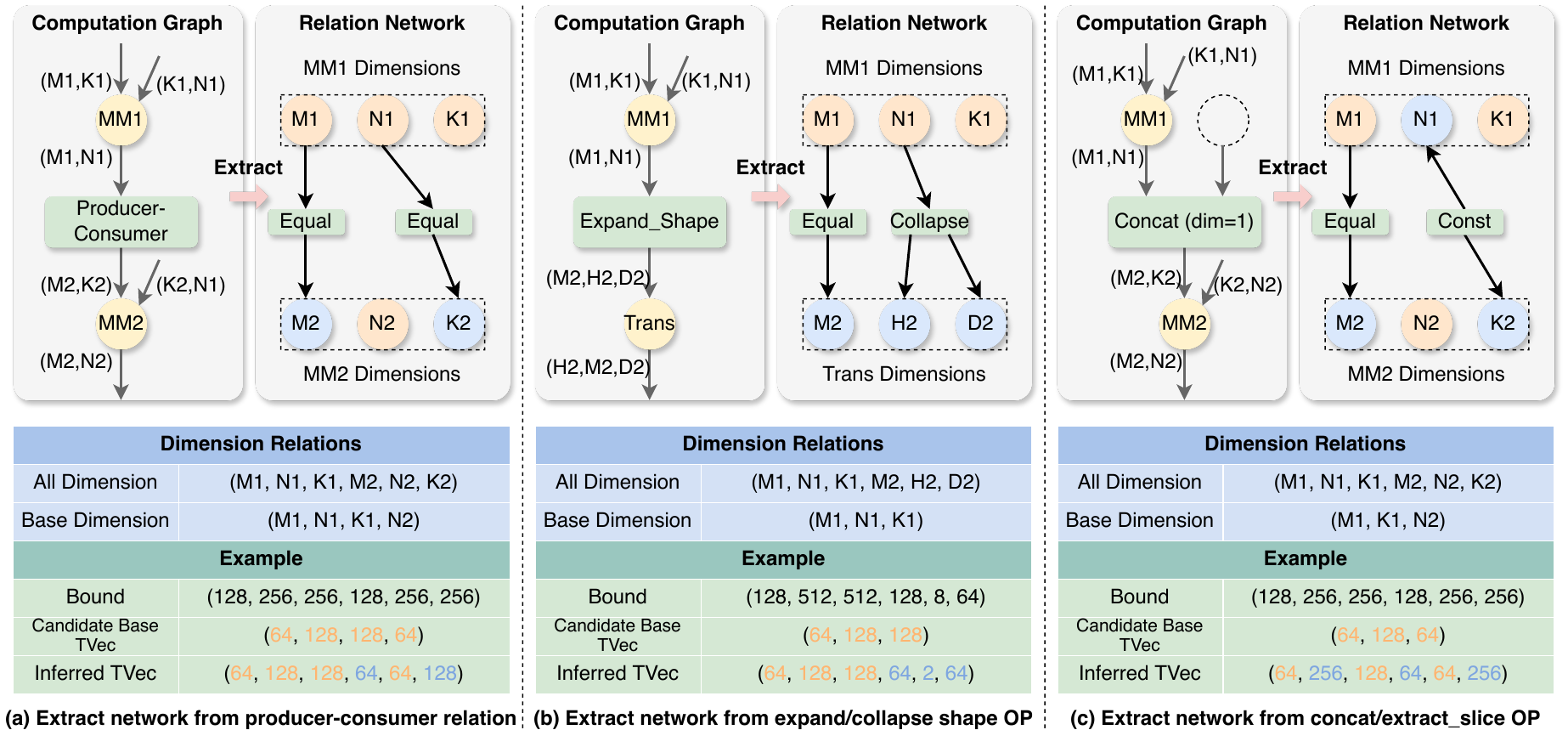} 
    \caption{\rev{An illustrative example of a dimension-relation network.} MM denotes a matrix multiplication operation, Trans denotes a transpose operation, and TVec denotes the tiling vector. Orange: base dimensions; blue: inferred dimensions.}
    \label{fig:dimension_relation_network}
\end{figure*}

\textit{\textbf{Dimension-Relation Network Representation.}}  
Within each cluster, due to the pipelined data transfer between operations through SRAM, there exist inherent constraints on the tiling sizes of producer and consumer operations.
For example, in Fig.~\ref{fig:dimension_relation_network}(a), if the producer MM1 generates data in \(64 \times 128\) tiles, it is desirable for the consumer MM2 to process data at the same granularity.
This implies strong dependencies among tiling sizes at corresponding dimensions across different operators. Therefore, instead of exhaustively enumerating all possible tiling sizes for each dimension, we can enumerate only a small set of representative dimensions and infer the rest, thereby reducing the search space greatly.

Based on this observation, we propose a \emph{dimension-relation network} representation, as shown in Fig. \ref{fig:dimension_relation_network}. It models the dependencies among operator dimensions as a directed graph, where nodes correspond to operation dimensions or relation nodes, and edges capture the dependency relationships. Each \textit{relation} node is associated with a forward function that infers the tiling size of output dimensions based on the input dimensions. We define three fundamental types of relations which are typical in real-world LLMs.

\underline{(1)} \textit{Equal Relation.}  
As shown in Fig.~\ref{fig:dimension_relation_network}(a), 
this common relation in pipelined execution arises from producer-consumer dependencies.
It enforces that the tiling size of one dimension must equal that of the corresponding dependent dimension. The forward function directly propagates the tiling size from the input dimension to the output dimension.

\underline{(2)} \textit{Collapse Relation.}  
It originates from 
\texttt{\seqsplit{tensor.collapse\_shape}} and \texttt{\seqsplit{\texttt{tensor.expand\_shape}}}.
These operations reinterpret tensors without modifying the underlying data.
As illustrated in Fig.~\ref{fig:dimension_relation_network}(b), an expand operation decomposes a dimension \(N_1\) into two dimensions \(H_2\) and \(D_2\).
Under this relation, the tiling sizes of the decomposed dimensions must satisfy a consistency constraint such that their product equals the original tiling size. Moreover, to ensure contiguous memory access, we adopt a low-to-high dimension allocation strategy, i.e., higher dimensions can only be tiled once lower dimensions reach their bounds. 
As shown in Fig.~\ref{fig:dimension_relation_network}(b), for \(N_1\), \(H_2\), and \(D_2\), a tiling configuration \((128, 2, 64)\) is valid because \(D_2\) reaches its bound, leading to contiguous access. In contrast, \((128, 4, 32)\) is a invalid configuration.

\underline{(3)} \textit{Constant Relation.}  
As shown in Fig.~\ref{fig:dimension_relation_network}(c), it is introduced by operations such as \texttt{concat} and \texttt{extract\_slice}, which either merge tensors along a dimension or extract slices. To ensure correctness and on-chip execution, the involved dimension and all lower dimensions must be fully tiled (i.e., equal to their bounds).

The relation network is constructed by traversing operations in a cluster in topological order. For each operation, \name{} examines its inputs and establish the corresponding dimension relations.
If an input is the direct output of a preceding operation, \name{} adds equal relations between the corresponding dimensions. If it is produced by \texttt{\seqsplit{collapse\_shape}} and \texttt{\seqsplit{\texttt{expand\_shape}}} operations, \name{} adds collapse relations for the affected dimensions and equality relations for the others. 
If the input is generated by \texttt{concat} and \texttt{extract\_slice} operations, \name{} assigns constant relations to the involved dimensions by fixing them to their bounds, while preserving equality relations for the remaining dimensions.
After the traversal, the dimension-relation network is obtained, as shown in Fig.~\ref{fig:dimension_relation_network}(a)--(c).



After constructing the relation network, we identify the base dimensions that cannot be inferred from others. We then enumerate tiling configurations only for these base dimensions and propagate each candidate through the network to infer the remaining tiling sizes, yielding a complete tiling vector for performance evaluation.

\begin{figure*}[]
    \centering
    \includegraphics[width=\textwidth]{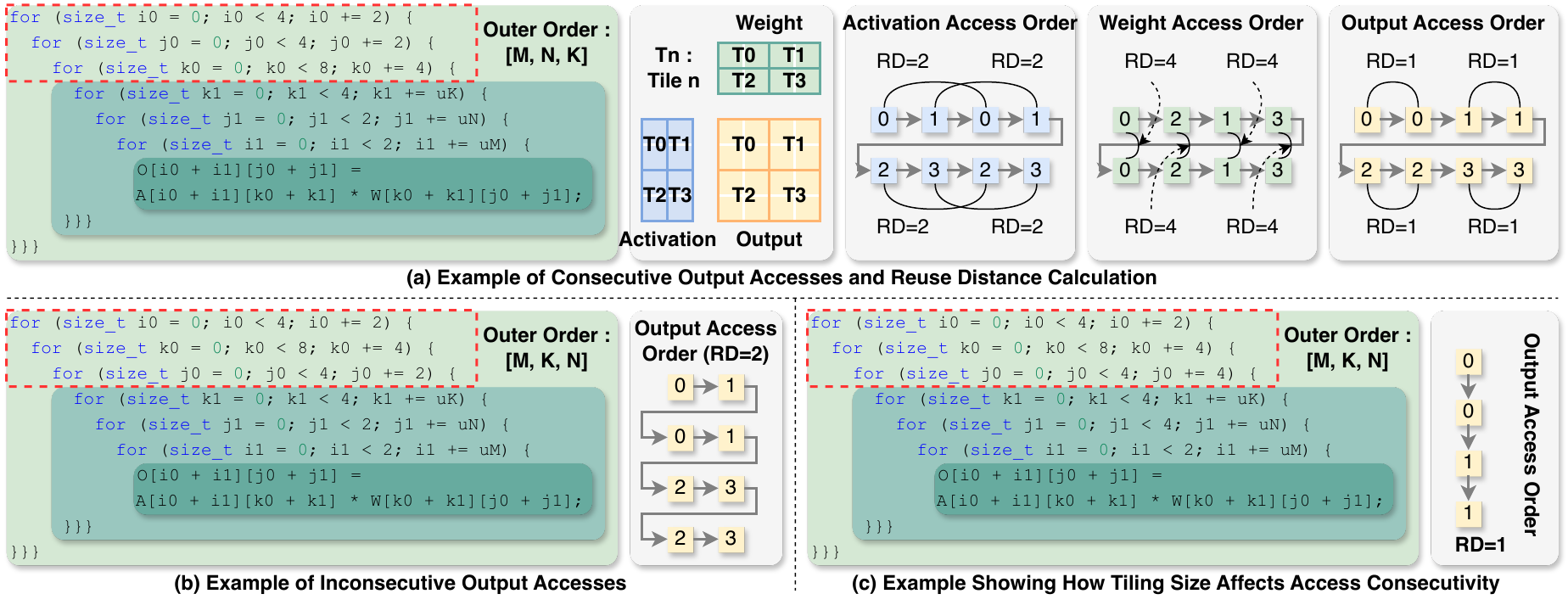} 
     \caption{\rev{Examples of different tensor access orders. RD denotes reuse distance.}}

    \label{fig:ordering_problem}
\end{figure*}


\textit{\textbf{Equivalence-Based Ordering Cache.}}
\rev{The outer-order vector is another critical L2 parameter, as it determines memory access patterns and thus DRAM efficiency. As with tiling sizes, operation ordering within a cluster is constrained. To enable direct tensor forwarding between producer and consumer, without buffering or reordering, the access order for a tensor must satisfy two conditions. \underline{First}, the reuse distance of each tile must be 1. Figure~\ref{fig:ordering_problem}(a) illustrates an example of the access order for each operand. With reuse distances of 2 and 1 for the activation and output tensors, respectively, the output tensor can be directly pipelined to the next operation, whereas the activation tensor cannot be forwarded directly from the previous operation. 
\underline{Second}, the producer and consumer must access the tensor in the same order. For example, an operation that directly consumes the output in Figure~\ref{fig:ordering_problem}(a) must access the output tiles in the same order in which they are produced, i.e., Tile0, Tile0, Tile1, Tile1, $\ldots$.
}

\rev{
The tiling size further complicates this constraint, as it can alter the access consecutivity of tensors.
Taking Figures~\ref{fig:ordering_problem}(b) and~(c) as examples, both cases follow the outer order $[M,K,N]$. However, different tiling sizes lead to different output access patterns: Figure~\ref{fig:ordering_problem}(b) is non-consecutive, whereas Figure~\ref{fig:ordering_problem}(c) is consecutive.
Thus, we must generate all valid outer-order vectors for each candidate tiling size. 
Although this can be achieved by performing a depth-first search (DFS),
the search space grows exponentially with both the cluster size and the number of candidate tiling vectors, making exhaustive enumeration impractical.
}

\rev{Our key observation is that different tiling configurations can fall into the same \emph{equivalence class} when searching for valid outer-loop orders, thereby significantly reducing the search cost. 
Consider the tiling size $(2,2,4)$ in Figure~\ref{fig:ordering_problem}(a) and (b). The only outer-loop orders that preserve consecutive output accesses are $(M,N,K)$ and $(N,M,K)$. If the tiling size is changed to $(2,2,2)$, the set of valid outer-loop orders remains unchanged. More generally, any tiling configuration that does not fully tile any dimension has the same set of valid outer-loop orders. We therefore group such tiling configurations into the same \emph{equivalence class}.
}

\rev{Based on this observation, we define a \emph{dimension mask} to indicate whether each dimension is fully tiled. All tiling configurations sharing the same dimension mask belong to the same equivalence class and thus share identical candidate order sets. Leveraging this property, we propose an equivalence-based ordering cache implemented as a hash map. During L2 parameter enumeration, for each tiling configuration, we compute its dimension mask (encoded as a 64-bit integer) and check whether the corresponding equivalence class has already been explored. If so, we directly reuse the cached candidate order set; otherwise, we perform DFS-based enumeration and 
store the results in the cache.}


\subsubsection{Solver-based Search for L1 Parameters.}  
The \emph{L1 parameters} of each operation are independent and can thus be optimized separately under shared system constraints. We model the execution of a cluster as a fully pipelined dataflow system, where each operation produces one output tile per iteration. The steady-state pipeline latency is determined by the slowest stage. Accordingly, we first optimize the per-operation parallelism \(p_i\) to minimize the overall pipeline latency. Let \(f_i\) denote the computational cost (in floating-point operations) required to produce one output tile of operation \(i\). 
The per-tile latency of operation \(i\) is \(f_i / p_i\), yielding the minimax optimization objective:
\(
\min_{\{p_i\}} \; \max_i \left(\frac{f_i}{p_i}\right),
\)
subject to global resource constraints :
\(
\sum_i n_{ij} \leq N_j.
\)

Here, \(n_{ij}\) denotes the amount of resource type \(j\) required by operation \(i\), and \(N_j\) is the corresponding available resource budget. By introducing an auxiliary variable \(t\), the problem can be equivalently reformulated as:
\[
\min_{t,\{p_i\}} \; t \quad \text{s.t.} \quad \frac{f_i}{p_i} \leq t,\ \forall i,\quad \sum_i n_{ij} \leq N_j,\ \forall j,
\]

which yields a standard form amenable to linear programming. This formulation can be efficiently solved using off-the-shelf linear programming solvers such as GLPK \cite{GLPK}.

\begin{figure*}[h!]
    \centering
    \includegraphics[width=\textwidth]{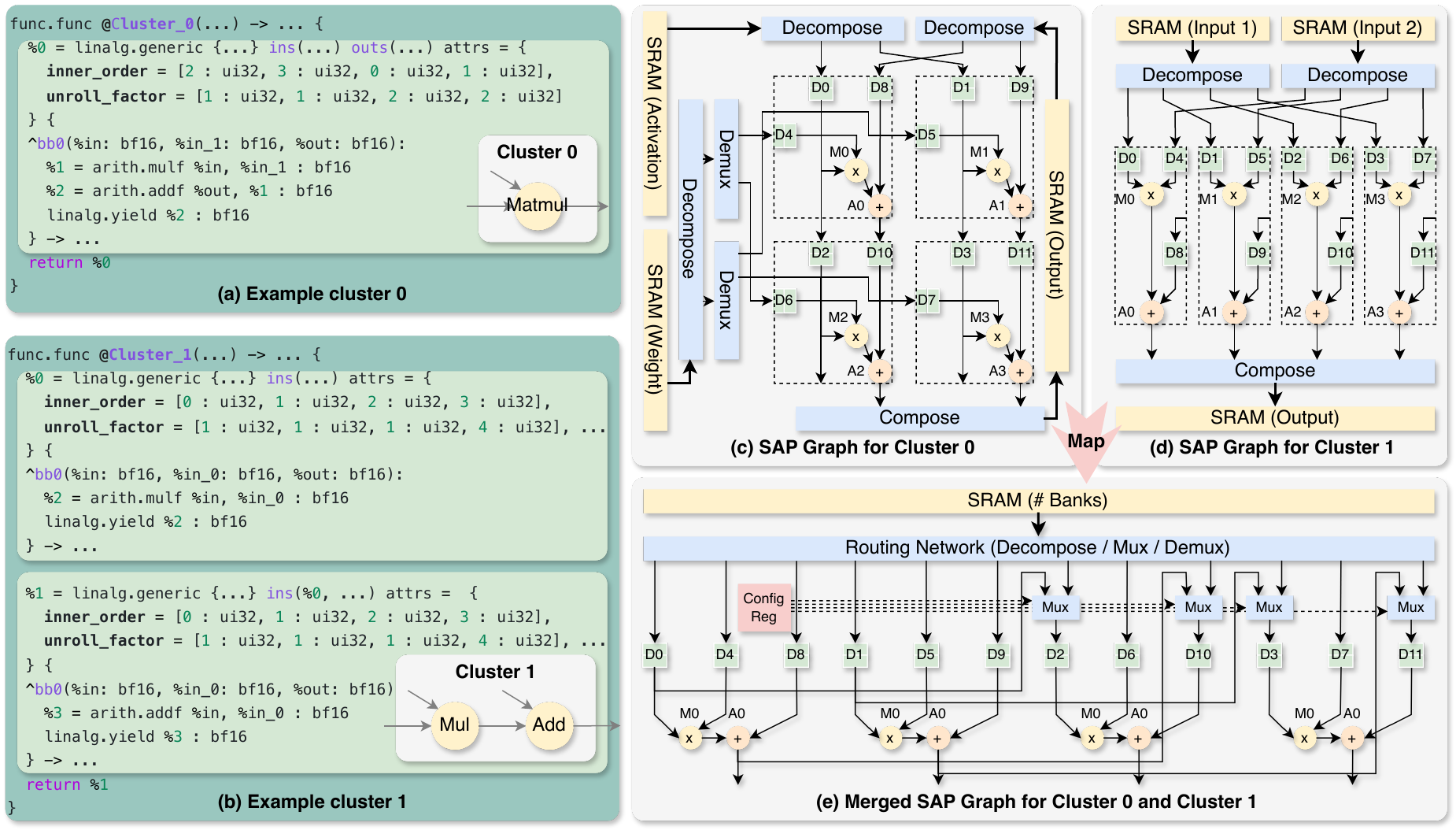} 
    \caption{\rev{Example of generated primitive graph. $D_n$ denotes DataNode $n$, $M_n$ denotes Multiplier $n$, and $A_n$ denotes Adder $n$.}}
    \label{fig:primitive_graph}
\end{figure*}

After obtaining the optimal parallelism configuration \(\{p_i^\star\}\), \name{} maps it to loop unrolling factors over 
dimensions. We restrict unrolling to at most two dimensions to match the 2D PE array.
We prioritize dimensions that preserve contiguous memory accesses for each operand tensor, improving spatial locality and
memory bandwidth utilization. The selected unroll dimensions are then placed at the outer loop levels of the L1 memory hierarchy to maximize data reuse across iterations.

%% file: section/backend.tex
\section{Hardware Generation}
\label{sec:back_end}

In this section, we introduce the abstraction termed spatial architecture primitives (SAPs) and describe how high-level representations
are progressively lowered to SAPs, ultimately generating binary and RTL code.


\subsection{Spatial Architecture Primitives}

Spatial architecture primitives (SAPs) are proposed to address the difficulty of generating hardware for multi-operation pipelined clusters with heterogeneous PEs and flexible topologies. The key idea is to model accelerators as directed cyclic graphs, where nodes represent storage, computation, or routing components, and edges represent dataflow dependencies.

To enable automated graph construction, we define four classes of primitives. For edges, \texttt{DataPath} represents connections between source and destination ports. For nodes, we define three categories: data, routing, and arithmetic primitives. Data primitives include FIFO- and SRAM-based storage components. Routing primitives handle data movement, such as \texttt{Decompose}, \texttt{Compose}, \texttt{Mux}, and \texttt{Demux}, by transforming data stream formats or directing data paths. Arithmetic primitives represent computation operations. All SAPs are summarized in Appendix~\ref{app:sap}.

\subsection{SAP Graph Construction}
\label{sec:sap_graph}

\rev{Taking scheduled clusters as input, the \name{} backend constructs a SAP graph to represent a reconfigurable architecture that supports all required dataflows of clusters. 
The construction process consists of three stages.}

\rev{\textit{\textbf{Primitive allocation.}}
First, based on the hardware resource configurations specified in the primitive library, we pre-instantiate storage and arithmetic primitives and organize them into a resource pool. 
For each cluster, we allocate arithmetic and DataNode resources to each \texttt{linalg} operation based on its unrolling factors, and SRAM to each producer--consumer relation based on its tiling size to enable data pipelining.
This design maps resources required by different dataflows onto a fixed set of physical primitives. 
As illustrated in Figures~\ref{fig:primitive_graph}(c) and (d), two different dataflows physically share the same set of DataNodes ($D0$--$D11$) and arithmetic primitives ($M0$--$M3$ and $A0$--$A4$).}

\rev{\textit{\textbf{Datapath construction.}}
After allocation, each \texttt{linalg} operation is equipped with a set of DataNodes, arithmetic primitives, and SRAM primitives. We organize the DataNodes and arithmetic primitives as a 2D array according to the unroll factors, while organizing the SRAM primitives as 1D vectors. The datapath construction stage connects these primitives by establishing the required data paths.}

\rev{This process consists of three sub-steps. \underline{(1)} \textit{Routing data from SRAM to DataNodes.} Since SRAM and DataNode array shapes are typically incompatible, we reshape and distribute SRAM data to the DataNodes using routing and datapath primitives. In Figures~\ref{fig:primitive_graph}(a) and (c), a \textit{Decompose} primitive partitions the activation input into \(1\times2\) blocks. With shared input dependencies, datapath primitives connecting adjacent PEs broadcast each element along the row to form the \(2\times2\) input required by the DataNode array. For weights, a \textit{Decompose} primitive similarly partitions the SRAM output, while a \textit{Demux} primitive preloads each \(2\times2\) weight block into different rows, as the iterations have no shared input dependencies. Finally, partial sums are initialized by reading a \(1\times2\) block; thus, only a \textit{Decompose} primitive is needed to directly feed the first row of output DataNodes.
}

\rev{\underline{(2)} \textit{Synthesizing the arithmetic datapath.} For the arithmetic datapath, we analyze the data dependencies within the \texttt{linalg} body and connect the corresponding DataNodes and arithmetic primitives accordingly.}


\rev{\underline{(3)} \textit{Routing data back to SRAM.}
Similar to the input path, the DataNode array and output SRAM may have incompatible shapes. We therefore insert routing primitives based on output dependencies. In Figures~\ref{fig:primitive_graph}(a) and (c), the \(1\times2\) output blocks and \(2\times2\) DataNode array imply a row-wise reduction, so datapath primitives forward partial sums along the row. A \textit{Compose} primitive then packs the outputs onto the wide SRAM bus for writing.
}




\rev{\textit{\textbf{ Datapath reorganization.}}
Since each \texttt{linalg} operation independently connects its storage and arithmetic primitives, the resulting datapaths may contain redundant primitives and lack support for reconfiguration. We therefore further reorganize the datapath to eliminate redundancy and enable reconfigurability. 
\underline{First}, we eliminate redundancy. When multiple instances of the same primitive connect to a single port, we reorganize the datapath to share one instance. A typical example is eliminating redundant \textit{Decompose} primitives connected to the same SRAM port. For \texttt{linalg} operations with the same unroll factor and inner order, we similarly eliminate intermediate SRAM and directly forward results. For example, in Cluster 1 of Figures~\ref{fig:primitive_graph}(b) and (d), the matching configurations allow the Adder output to be directly forwarded to the Multiplier.
\underline{Second}, we introduce reconfigurability by inserting a \textit{Mux} whenever a primitive has multiple possible input sources, allowing the source to be selected at runtime through configuration. As illustrated in Figure~\ref{fig:primitive_graph}(e), configuring these \textit{Mux} primitives with configuration registers enables the accelerator to execute either of the dataflows shown in Figure~\ref{fig:primitive_graph}(c) and (d). This reconfigurability incurs only minor overhead (Sec.~\ref{sec:ablation}).}

\subsection{Binary and RTL Code Generation}

\rev{
After constructing the SAP graph, 
\name{} generates the binary and RTL code.
We map each cluster to a binary which encodes parameters including tiling, unrolling, loop orders, and multiplexer control bits, 
while runtime fields specify loop bounds and memory addresses.
These binaries are stored in the dispatch table for runtime execution.
Meanwhile, the RTL code is generated by replacing each primitive in the SAP graph with its RTL template or hardware macro.
}

%% file: section/evaluation.tex
\section{Evaluation}

\subsection{Experimental Setup}

\textit{\textbf{Workloads.}} We select Llama3-8B \cite{LLAMA3}, Qwen3-8B \cite{QWEN3}, Gemma-7B \cite{GEMMA}, Llama3-70B \cite{LLAMA3}, and Qwen3-235B-A22B (balanced routing, abbreviated as Qwen3-MoE) \cite{QWEN3MOE} as representative LLM workloads. Due to space constraints, we report the Llama3-8B results in Appendix~\ref{app:llama70b}.

\rev{\textit{\textbf{GPU and NPU Baselines.}} 
We use the NVIDIA Orin NX \cite{ORINNX} and NVIDIA A100 GPU \cite{A100} as baselines. 
We also compare against the Ascend 910B3 NPU \cite{ASCEND}, with BF16 peak throughput and memory bandwidth comparable to the A100.
We use implementations from Transformers library \cite{TRANSFORMERS}, with native PyTorch \cite{TORCH} and Torch-NPU \cite{TORCHNPU} backends for NVIDIA and Ascend devices, respectively, leveraging optimized device-specific kernels.
To obtain accurate latency measurements, we explicitly synchronize device execution using \texttt{torch.cuda.synchronize()} and \texttt{torch.npu.synchro-} \texttt{nize()} before recording execution time. We measure power consumption using \texttt{tegrastats}, \texttt{nvidia-smi}, and \texttt{npu-smi} for the Orin NX, A100, and Ascend 910B3, respectively.}

\textit{\textbf{Accelerator Generator Baselines.} }
We compare \name{} with two accelerator generators: Gemmini \cite{GEMMINI} and LEGO \cite{LEGO}.
For Gemmini, we consider an architecture consisting of a systolic array and a vector processing unit operating under output-stationary and weight-stationary dataflows, denoted as GEMMINI-OS and GEMMINI-WS, respectively. For LEGO, due to its flexibility, we instantiate an accelerator composed of a systolic array and a vector processing unit with a configurable dataflow that can dynamically switch among input-, weight-, and output-stationary modes.

\begin{figure*}[h!]
    \centering
    \begin{minipage}{0.495\linewidth}
        \centering
        \includegraphics[width=\linewidth]{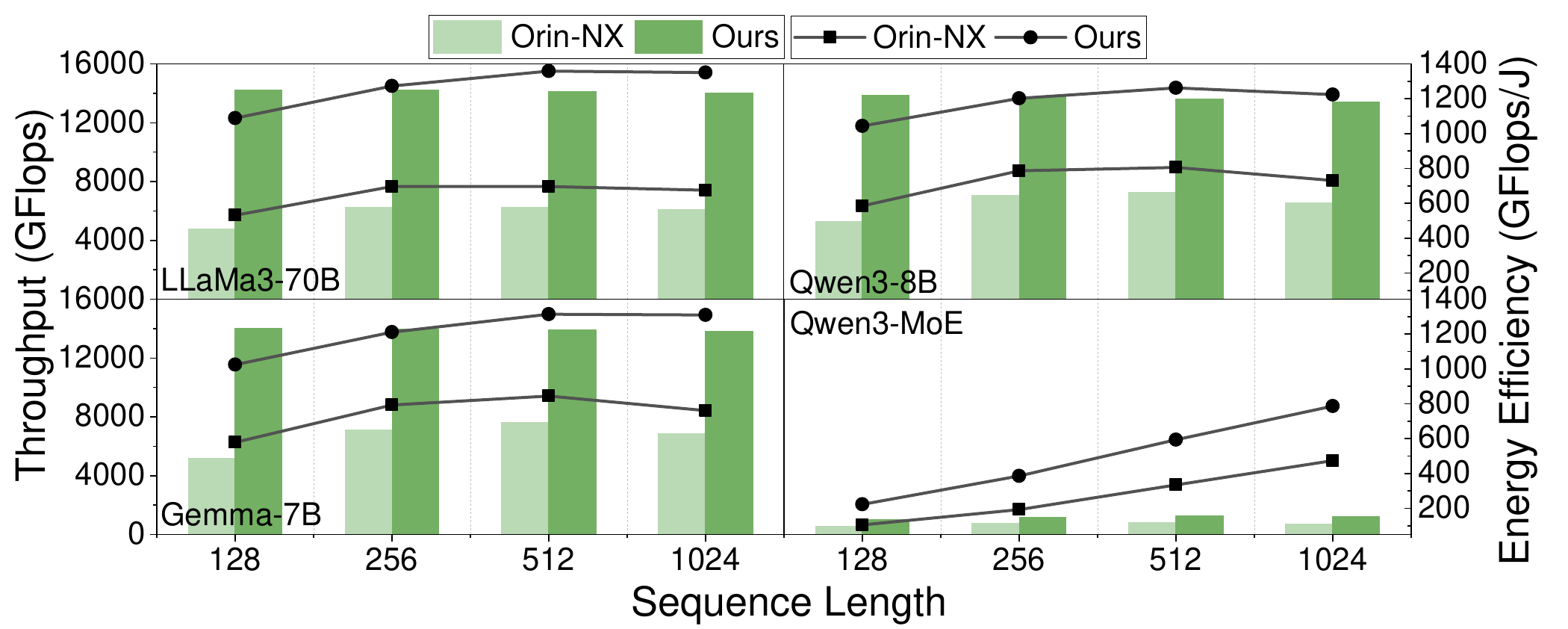}
        \subcaption{Edge configuration + Prefill (Batch=1).}
    \end{minipage}
    \hfill
    \begin{minipage}{0.495\linewidth}
        \centering
        \includegraphics[width=\linewidth]{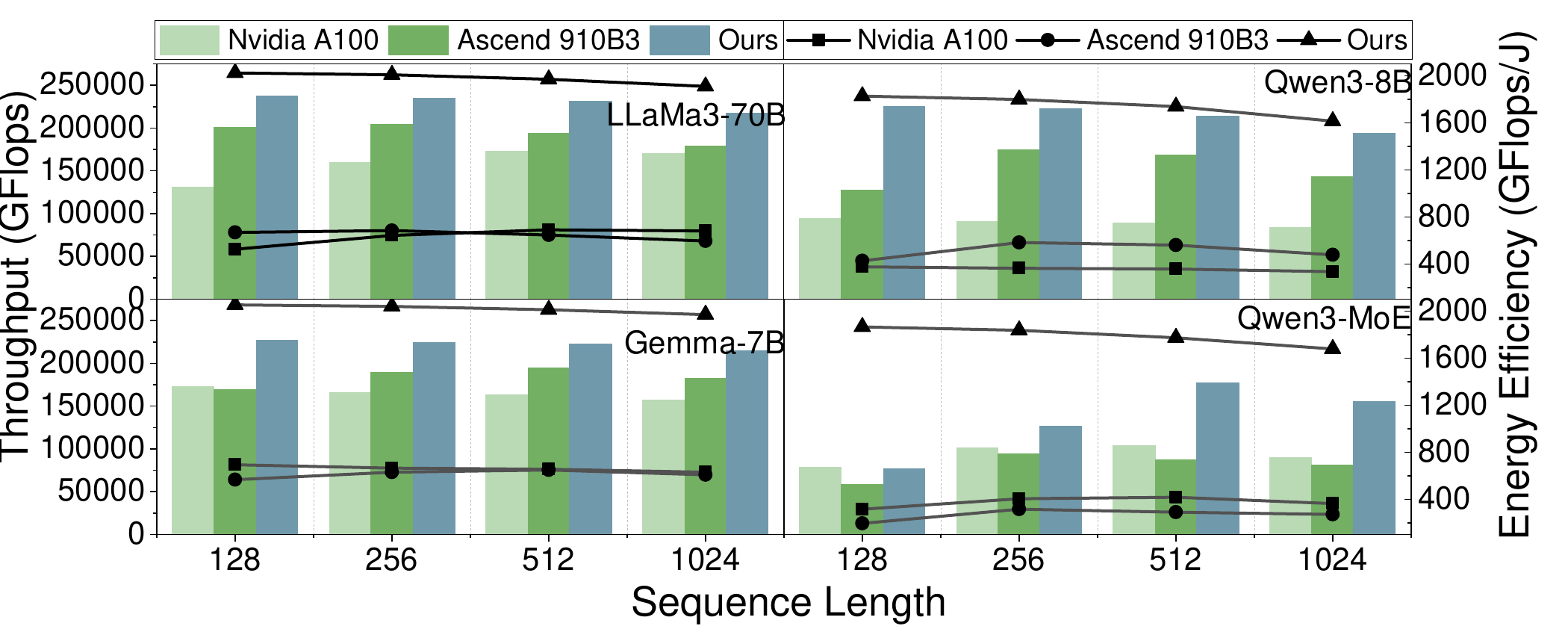}
        \subcaption{Server configuration + Prefill (Batch=8).}
    \end{minipage}
    \begin{minipage}{0.495\linewidth}
        \centering
        \includegraphics[width=\linewidth]{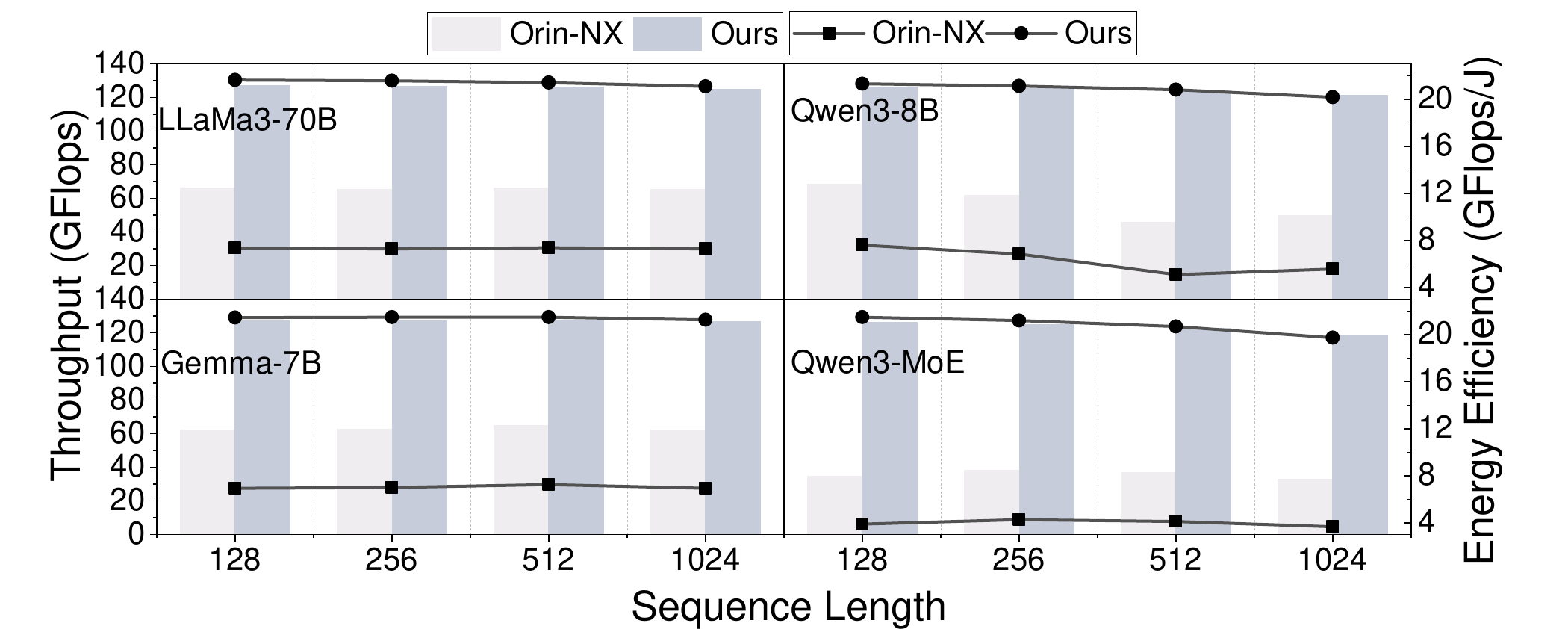}
        \subcaption{Edge configuration + Decode (Batch=1).}
    \end{minipage}
    \hfill
    \begin{minipage}{0.495\linewidth}
        \centering
        \includegraphics[width=\linewidth]{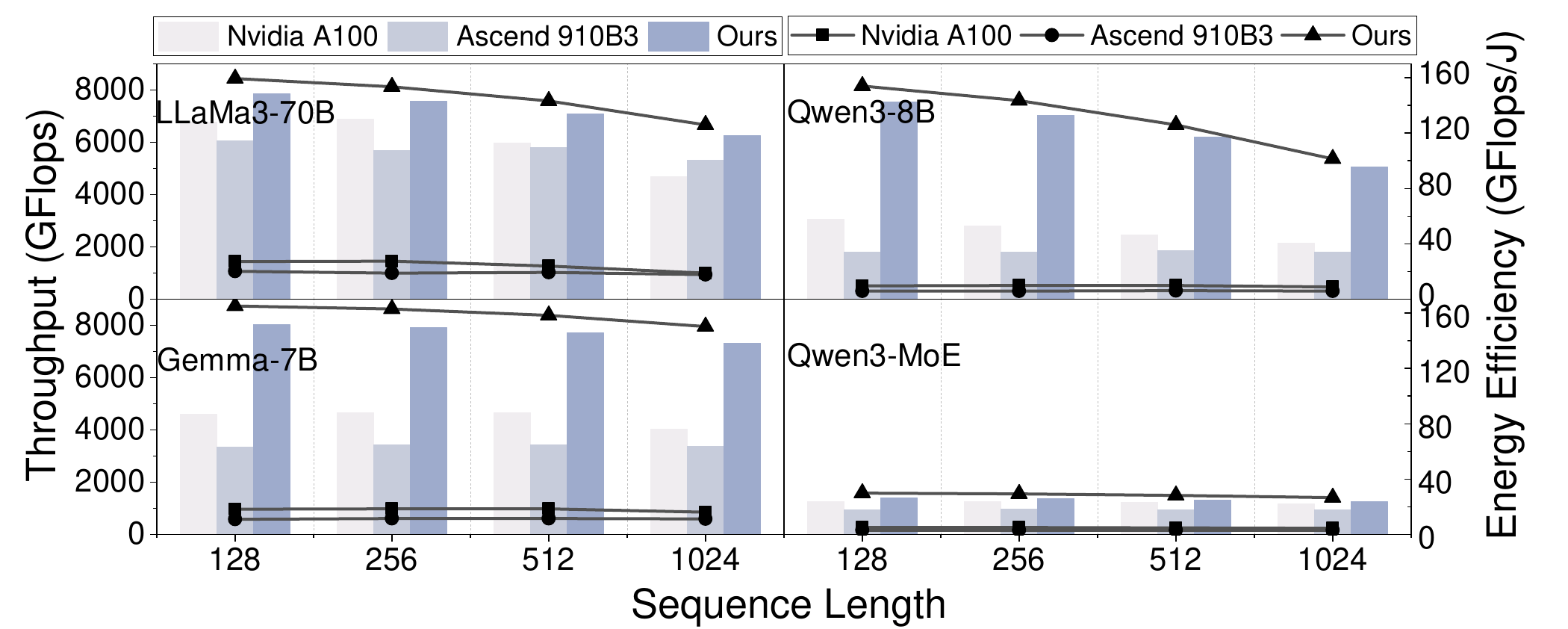}
        \subcaption{Server configuration + Decode (Batch=8).}
    \end{minipage}
    \caption{\rev{Throughput and energy-efficiency comparison across GPUs and commercial accelerators. Bars show throughput (left y-axis), and lines show energy efficiency (right y-axis).}}

    \label{fig:tp_ee_gpu}
\end{figure*}

We model the baselines using Timeloop and Accelergy \cite{TIMELOOP}, which support accelerator architectures and SRAM/DRAM modeling.
For each architecture, we perform mapping exploration using Timeloop and select the mapping that minimizes EDP under the same optimization criteria and search cost, reporting corresponding performance results. 
Since Timeloop does not support modeling multiple pipelined operations, we extract latency from our performance model for generated accelerators and combine it with energy estimates from Accelergy to obtain the final metrics.



All experiments are conducted on a system equipped with an Intel Xeon Gold 5218 CPU. 
We use  edge/server configurations for accelerator-generator comparisons and the largest valid PE configuration with BF16 peak throughput no greater than the target device for commercial-devices comparisons.
All TranSAGE configurations run at 1 GHz using a 45 nm technology model, with parameters summarized in Appendix B. 
We use DeepScaleTool \cite{DEEPSCALETOOL} to scale energy estimates to 8 nm for Orin NX and 7 nm for A100 and Ascend 910B3.




\subsection{Comparison with GPU and NPU}

\rev{\textit{\textbf{Throughput. }}
As shown in Figure~\ref{fig:tp_ee_gpu}, for prefill, \name{} achieves average throughput improvements of 2.05$\times$, 1.65$\times$, and 1.42$\times$ over the Orin NX, A100, and Ascend 910B3 NPU, respectively.
For dense models, including LLaMA3-70B, Qwen3-8B, and Gemma-7B, the improvement mainly comes from pipelining auxiliary operations, which further increases overall throughput even when the matrix multiplications (MM) already achieve relatively high utilization. For Qwen3-MoE, its MoE structure results in low compute density in the FFN operations, allowing \name{} to further benefit from fusing short-sequence MM operations.
For the memory-bound decode workload, \name{} achieves average throughput improvements of 2.41$\times$, 1.62$\times$, and 2.12$\times$ over the Orin NX, A100, and Ascend 910B3, respectively, demonstrating the benefits of operation fusion for memory-bound workloads.}


\rev{\textit{\textbf{Energy efficiency.} }
\name{} achieves average energy-efficiency improvements of 1.78$\times$, 3.46$\times$, and 3.69$\times$ over Orin NX, A100, and 910B3, respectively, on prefill, demonstrating the energy-efficiency advantage of our specialized architecture.
For the decode workload, where memory accesses dominate energy consumption, \name{} achieves average energy-efficiency improvements of 3.65$\times$ over the Orin NX and 8.57$\times$ and 13.04$\times$ over the A100 and 910B3, respectively. These gains stem primarily from the reduction in memory traffic enabled by operation fusion.}

\begin{figure*}[]
    \centering
    \begin{minipage}{0.495\linewidth}
        \centering
        \includegraphics[width=\linewidth]{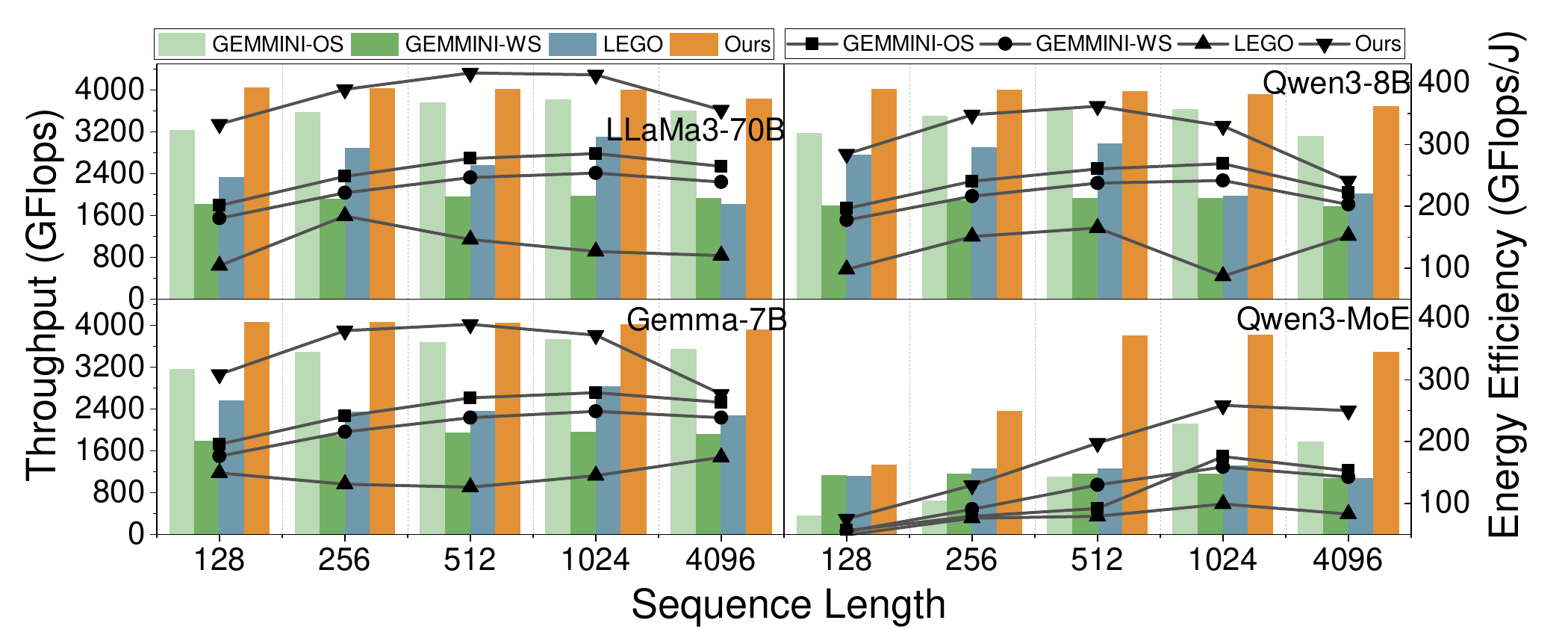}
        \subcaption{Edge configuration + Prefill (Batch=1).}
    \end{minipage}
    \hfill
    \begin{minipage}{0.495\linewidth}
        \centering
        \includegraphics[width=\linewidth]{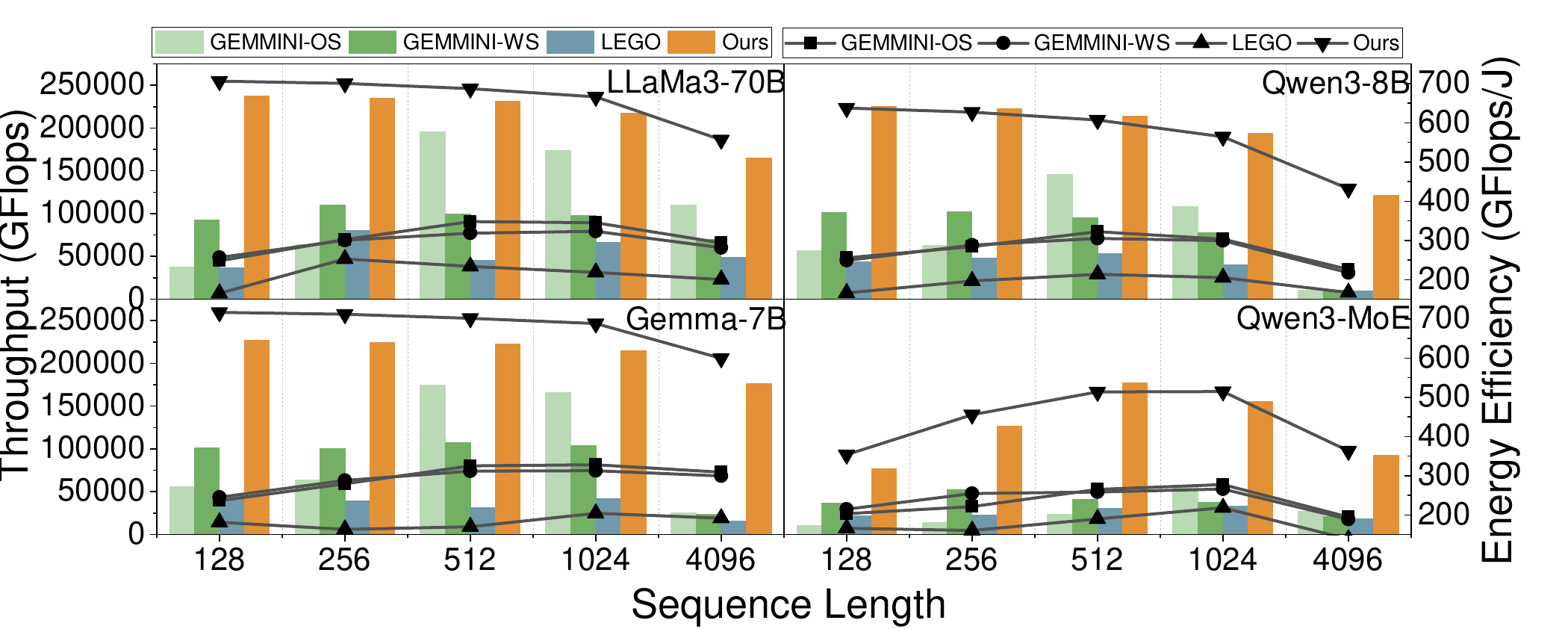}
        \subcaption{Server configuration + Prefill (Batch=8).}
    \end{minipage}
    \begin{minipage}{0.495\linewidth}
        \centering
        \includegraphics[width=\linewidth]{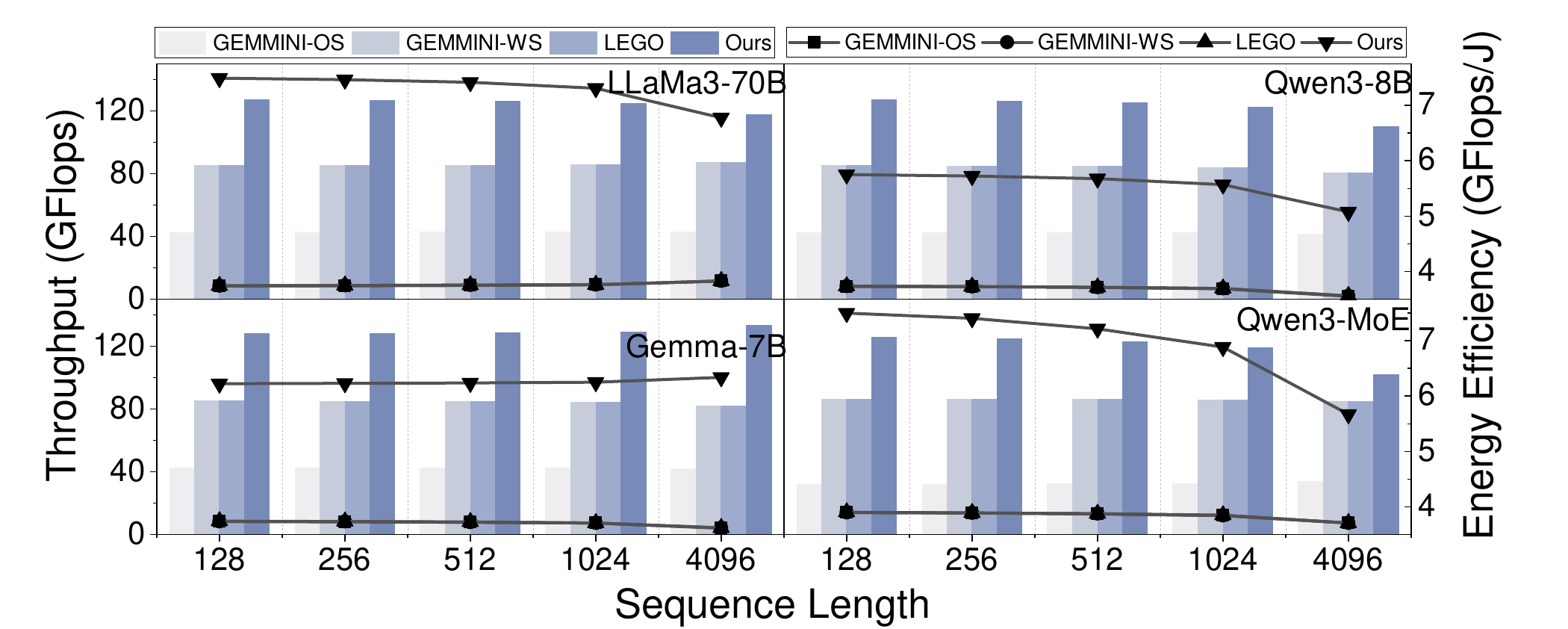}
        \subcaption{Edge configuration + Decode (Batch=1).}
    \end{minipage}
    \hfill
    \begin{minipage}{0.495\linewidth}
        \centering
        \includegraphics[width=\linewidth]{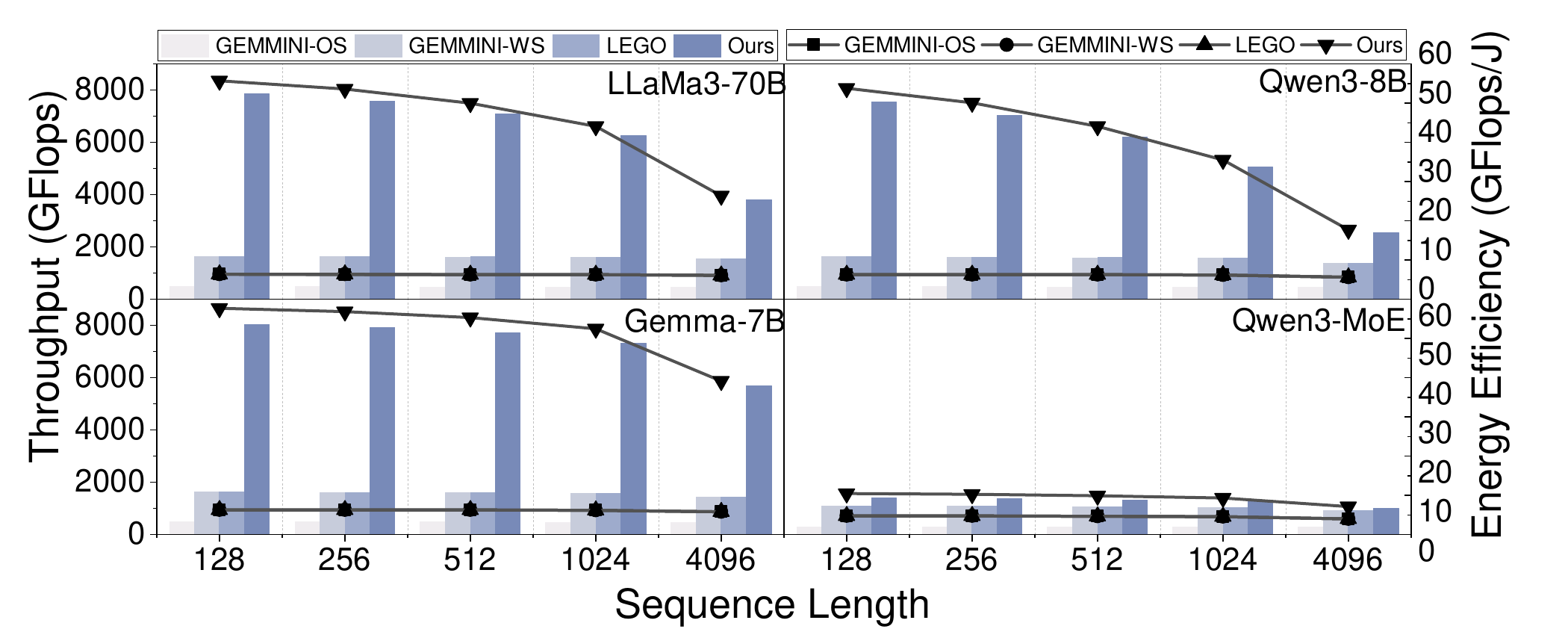}
        \subcaption{Server configuration + Decode (Batch=8).}
    \end{minipage}
    \caption{\rev{Throughput and energy-efficiency comparison with state-of-the-art accelerator generators. Bars show throughput (left y-axis), and lines show energy efficiency (right y-axis).}}

    \label{fig:tp_ee}
\end{figure*}

\subsection{Comparison with SOTA Accelerator Generators}


\textit{\textbf{Throughput.}} We next compare \name{} with SOTA accelerator generators. As shown in Figure~\ref{fig:tp_ee}, for the prefill workload, \name{} achieves average throughput improvements of 1.58$\times$, 2.22$\times$, and 1.82$\times$ over GEMMINI-OS, GEMMINI-WS, and LEGO, respectively, under the edge configuration. Under the server configuration, \name{} achieves 4.14$\times$, 3.28$\times$, and 5.50$\times$ improvements over the same baselines. Similar to the GPU comparison, these gains mainly come from pipelining auxiliary operations and reducing memory accesses for memory-bound operations. Moreover, \name{} maintains consistent throughput improvements across different sequence lengths, demonstrating the effectiveness of its adaptive dataflow.
For the decode workload, \name{} further achieves average throughput improvements of 3.13$\times$, 1.46$\times$, and 1.46$\times$ over GEMMINI-OS, GEMMINI-WS, and LEGO, respectively, under the edge configuration. Under the server configuration, the improvements increase to 11.33$\times$, 3.37$\times$, and 3.37$\times$, respectively. 

\begin{figure*}[h!]
    \centering
    \begin{minipage}{0.49\linewidth}
        \centering
        \includegraphics[width=\linewidth]{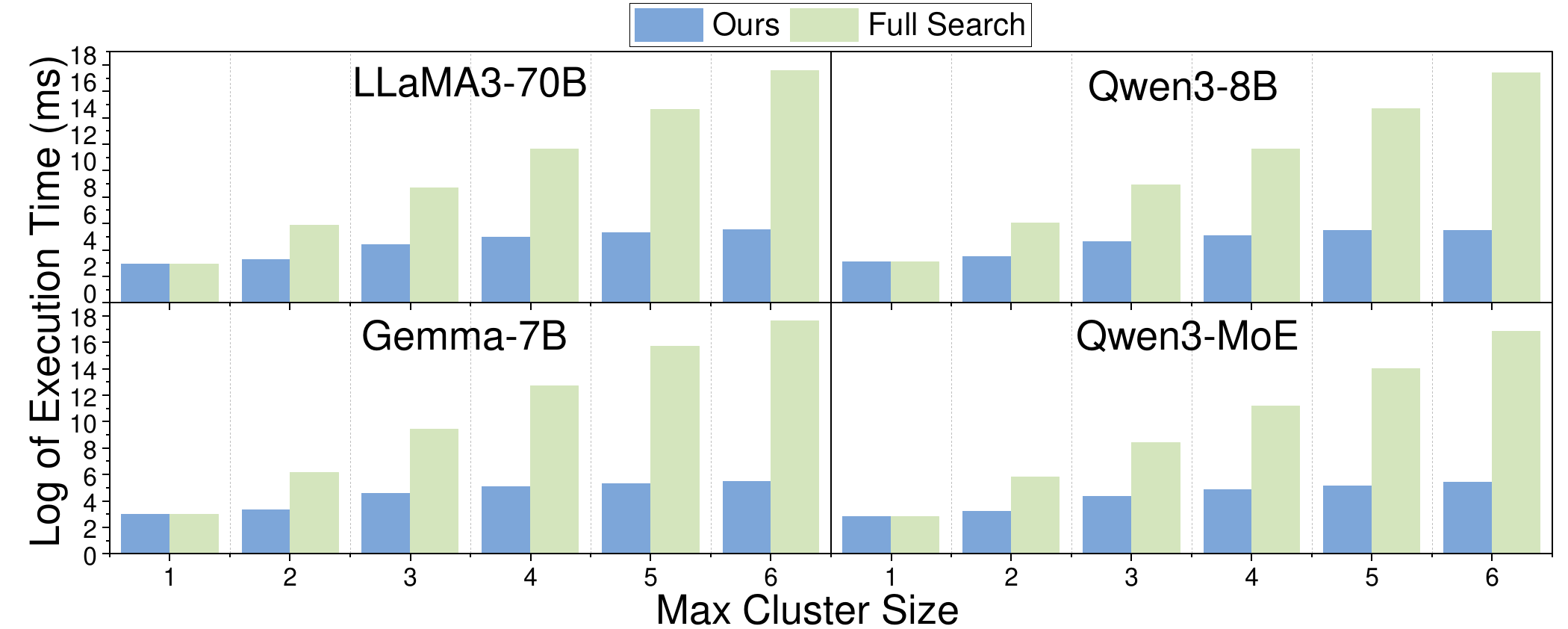}
        \subcaption{Execution time breakdown. }
    \end{minipage}
    \hfill
    \begin{minipage}{0.24\linewidth}
        \centering
        \includegraphics[width=\linewidth]{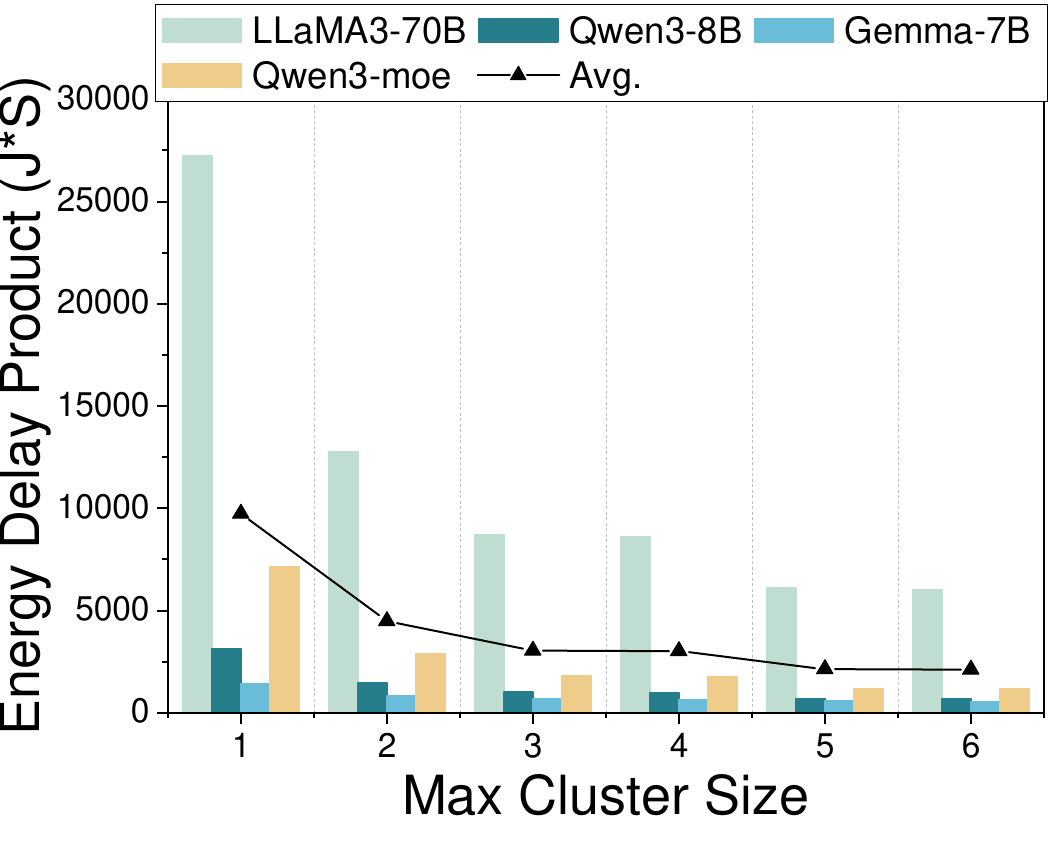}
        \subcaption{EDP breakdown.}
    \end{minipage}
    \hfill
    \begin{minipage}{0.24\linewidth}
        \centering
        \includegraphics[width=\linewidth]{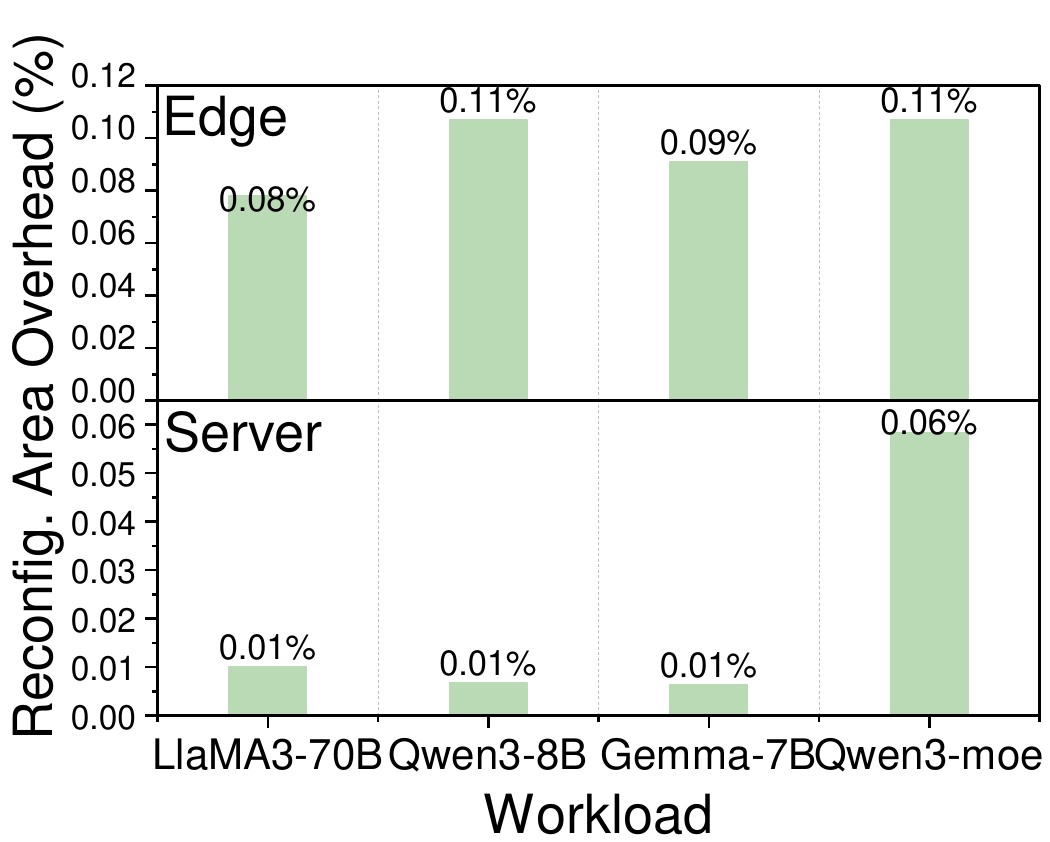}
        \subcaption{Area overhead breakdown.}
    \end{minipage}
    
    \caption{\rev{Ablation study of the maximum cluster size and reconfiguration area overhead.}}

    \label{fig:breakdown}
\end{figure*}

\rev{\textit{\textbf{Utilization.} }
Using the throughput reported in Figure \ref{fig:tp_ee}, we analyze effective BF16 arithmetic utilization.
\name{} achieves up to 96.15\% and 90.20\% utilization on the prefill workload under the edge and server configurations, respectively. This high utilization is enabled by the high compute density and the sharing of the same set of hardware resources across all clusters. In contrast, on the decode workload, \name{} achieves 3.16\% and 3.05\% utilization under the edge and server configurations, respectively, primarily due to the limited memory bandwidth.
}


\textit{\textbf{Energy efficiency.} }
Operation fusion also improves energy efficiency.
For the prefill workload, \name{} achieves average energy efficiency improvements of 1.46$\times$, 1.56$\times$, and 2.53$\times$ over GEMMINI-OS, GEMMINI-WS, and LEGO, respectively, under the edge configuration. It increases to 2.13$\times$, 2.15$\times$, and 3.13$\times$ under the server configuration. For the decode workload, \name{} achieves up to 1.73$\times$ and 6.07$\times$ energy efficiency improvement over the baselines under the edge and server configurations, respectively.

\subsection{Ablation Study}
\label{sec:ablation}



To further demonstrate the feasibility and performance of \name{}, we provide a breakdown analysis of the impact of the maximum cluster size, which limits the number of operations that can be grouped into a cluster, as well as the overhead introduced by reconfiguration.



\begin{figure*}[h!]
    \centering
    \includegraphics[width=\textwidth]{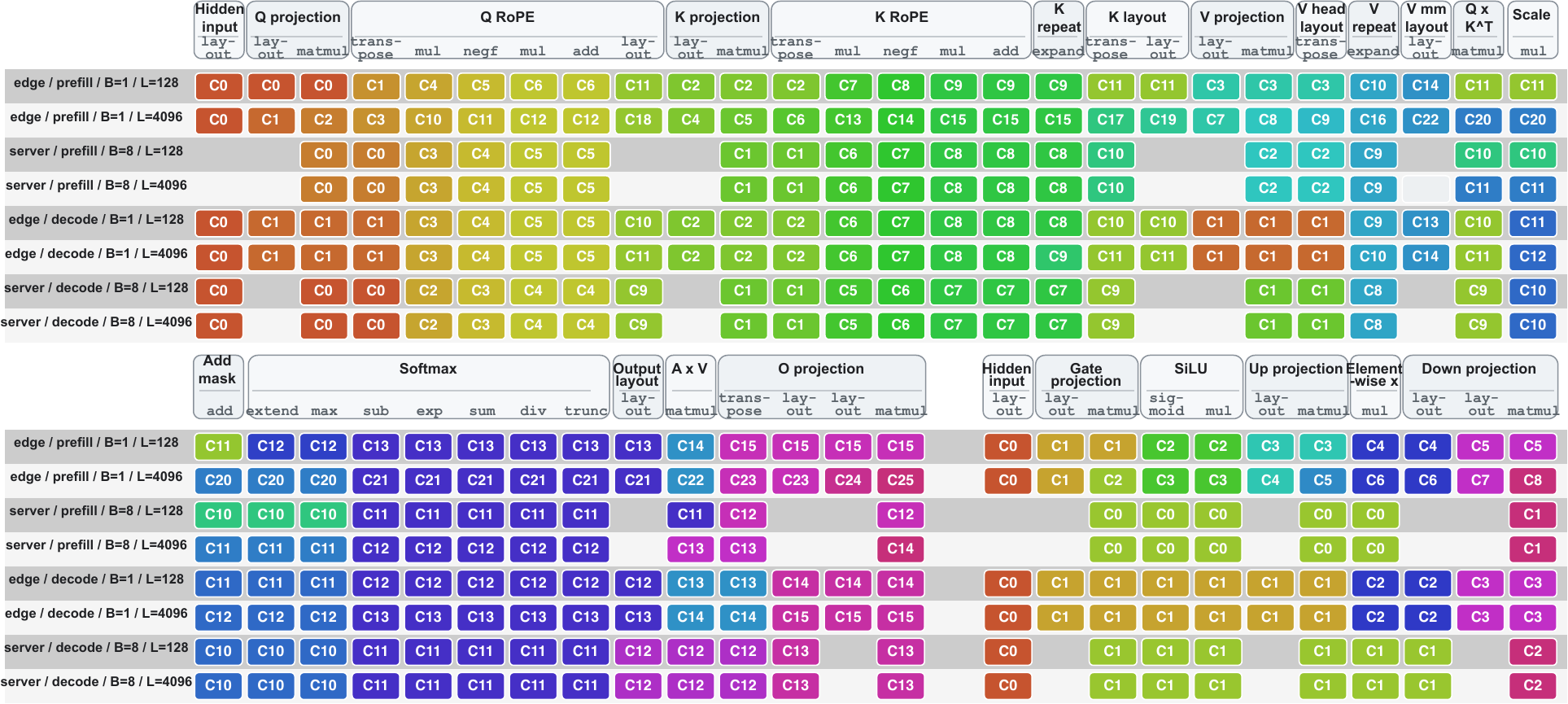} 
    \caption{\rev{Illustration of the resulting clusters from DSE across eight representative workloads. Each $C_n$, shown in a unique color, indicates that the corresponding operation is fused into cluster $n$ for the given workload.}}
    \label{fig:cluster_matrix}
\end{figure*}

\textit{\textbf{Execution Time.}} First, we evaluate how the maximum cluster size affects the execution time of \name{} in solving the scheduling problem for multi-operation clusters. As shown in Fig. \ref{fig:breakdown} (a), we compare the DSE time of \name{} with that of a full-search method which exhaustively enumerates all parameters in the prefill stage.
We observe that the execution time of full search grows exponentially with the maximum cluster size, becoming impractical once the cluster size exceeds 4. In contrast, \name{}, which relies on search-space pruning and decomposition, exhibits an almost linear increase in search time. Even with a maximum cluster size of 6, the search completes within tens of minutes, demonstrating the efficiency of our framework.

\textit{\textbf{EDP.}} We further evaluate how the maximum cluster size affects the EDP. Fig. \ref{fig:breakdown} (b) presents the EDP breakdown under varying maximum cluster sizes in the prefill stage with sequence length 4096. When the maximum cluster size is 1, i.e., without pipelining, all models exhibit relatively high EDP. As the maximum cluster size increases, EDP drops rapidly, further confirming the effectiveness of multi-operation pipelining. However, the improvement gradually saturates. This is because beyond a certain cluster size, pipelining can no longer effectively improve PE utilization; instead, operations begin to compete for shared PE and SRAM resources, leading to contention, lower utilization, and reduced data reuse. 


\rev{\textit{\textbf{Reconfiguration overhead.}}
As discussed in Section \ref{sec:sap_graph}, multiplexers are introduced to support dataflow reconfiguration. We therefore analyze the resulting area overhead. As shown in Figure \ref{fig:breakdown} (c), the area overhead due to reconfiguration is negligible across all evaluated models and configurations, reaching at most 0.11\%. This is because, for the same bit width, multiplexers occupy substantially less area than arithmetic PEs such as MAC units. Moreover, each model requires only a limited number of distinct dataflow configurations, further limiting the overall reconfiguration overhead.}

%% file: section/discussion.tex
\section{Discussion}

\subsection{Analysis of Fused Clusters}


\rev{
Fusion patterns largely determine topology flexibility and thus the resulting architectures.
To characterize which operations are suitable for fusion, Figure~\ref{fig:cluster_matrix} presents the resulting clusters for LLaMA-8B 
across representative edge/server, prefill/decode, and short/long-sequence workloads.
}

\rev{\textit{\textbf{Observations.}} Figure~\ref{fig:cluster_matrix} reveals two notable features of the fused clusters. \underline{First}, \textit{fusion patterns are workload- and resource-dependent}: under different configurations, the same model exhibits different fusion patterns. In general, edge configurations tend to produce more clusters than server configurations, while long-sequence workloads tend to result in more clusters than short-sequence workloads. This difference stems from variations in internal memory requirements and the available hardware resources.
\underline{Second}, \textit{effective fusion crosses conventional operator boundaries}. Operations that are typically treated as indivisible units in conventional designs, such as Softmax, RoPE and Gate projection, are often split across multiple clusters or fused with operations from other stages. 
These observations make it challenging to manually design an accelerator for each workload, highlighting the capability of \name{} to enable rapid design exploration across diverse workload and hardware configurations.} 


\rev{\textit{\textbf{Design insights.} }
Nevertheless, the generated partitions also reveal recurring principles that can guide future spatial-accelerator designs.
\underline{First}, 
\textit{fusing data-dependent operations that leverage complementary compute resources is generally beneficial.}
Such fusion can reduce memory accesses while improving resource utilization, making these operations ideal candidates for fusion. As shown in Figure~\ref{fig:cluster_matrix}, consecutive elementwise operations, such as Sub, Exp, Sum, Div, and Trunc in Softmax, as well as Mul and Add in RoPE, utilize different compute resources and can therefore be naturally fused to form an efficient pipeline. }

\rev{\underline{Second},
\textit{fusing data-dependent operations that compete for compute resources is beneficial only when they can form a balanced pipeline.}
More fusion is not necessarily better, especially when the fused operations compete for the same resources.
A typical example is Q/K/V projection. Q and K/V projections share the same input and thus have potential for fusion, but they are split into separate clusters under all prefill configurations, as shown in Figure~\ref{fig:cluster_matrix}.
This is because both operations consume substantial arithmetic resources during prefill, while group-query attention (GQA) results in different output dimensions for Q and K/V projections, leading to imbalanced workloads and resource underutilization, as discussed in Section~\ref{sec:motivation}, Pitfall 1.
In contrast, as shown in Figure~\ref{fig:cluster_matrix}, the gate and up projections in the FFN can be efficiently fused under server-prefill configurations. 
The reason is that these operations have identical output shapes and balanced computational demands, enabling even resource allocation and a balanced pipeline.
}


\subsection{Limitations and Future Directions}

\rev{\name{} has two limitations. 
First, its physical compute, storage, and interconnect resources are fixed after generation. Despite runtime reconfiguration, 
evolving model structures
may require hardware regeneration. 
Second, \name{} optimizes single-chip architectures. While it can be extended to multi-chip systems through data parallelism, it does not explore other parallelism strategies that account for inter-chip communication. Future work will aim at
providing greater flexibility in generated architectures, as well as communication-aware multi-chip DSE.}

%% file: section/conclusion.tex
\section{Conclusion}
\label{sec:conclusion}

We present \name{}, an MLIR-based framework for automatically generating spatial accelerators for large-scale transformers. By treating operation clusters as execution units, \name{} enables efficient pipelining and improves throughput and energy efficiency over GPUs, NPUs, and prior accelerator generators for both prefill and decoding.

%% file: section/appendix.tex
\clearpage
\appendix

\begin{table*}[!t]
\begin{minipage}{\textwidth}

\section{Spatial Architecture Primitives}
\label{app:sap}

\centering
\begin{tabularx}{\textwidth}{@{}llllX@{}}
\toprule
\textbf{Primitive} & \textbf{Op} & \textbf{Operand} &
\textbf{Results} & \textbf{Description} \\
\midrule
Link & DataPath & src, dest & / &
A point-to-point connection that transfers data between two ports. \\
\midrule
\multirow{2}{*}{Data}
& DataNode & / & din, dout &
A FIFO-based storage node that buffers streaming data. \\
& SramBank & / & write, waddr, din, read, raddr, dout &
An SRAM-based storage node supporting indexed reads and writes. \\
\midrule
\multirow{6}{*}{Routing}
& Mux & / & dins[...], dout &
Selects one input and forwards it to the output. \\
& DeMux & / & din, douts[...] &
Routes one input to one of multiple destinations. \\
& Broadcast & / & din, douts[...] &
Replicates input data to multiple output ports. \\
& Reduction & / & dins[...], dout &
Aggregates multiple inputs using an operation such as sum or max. \\
& Compose & / & dins[...], dout &
Combines multiple elements into a structured output. \\
& Decompose & / & din, douts[...] &
Splits structured input data into multiple elements. \\
\midrule
\multirow{8}{*}{Arithmetic}
& Mulf & / & lhs, rhs, result & Floating-point multiplication. \\
& Addf & / & lhs, rhs, result & Floating-point addition. \\
& Subf & / & lhs, rhs, result & Floating-point subtraction. \\
& Divf & / & lhs, rhs, result & Floating-point division. \\
& Maxf & / & lhs, rhs, result & Floating-point maximum. \\
& Negf & / & src, result & Floating-point negation. \\
& Truncf & / & src, result & Floating-point truncation. \\
& Extf & / & src, result & Floating-point extension. \\
\bottomrule
\end{tabularx}

\section{System Configurations}
\label{app:config}

\begin{threeparttable}
\begin{tabularx}{\textwidth}{
@{}l
>{\centering\arraybackslash}X
>{\centering\arraybackslash}X
>{\centering\arraybackslash}X
>{\centering\arraybackslash}X@{}
}
\toprule
\textbf{Resource} &
\textbf{Edge} &
\textbf{Orin NX matched} &
\textbf{Server} &
\textbf{A100/Ascend 910B3 matched} \\
\midrule
DRAM bandwidth & 128 GB/s & 128 GB/s & 1024 GB/s & 1024 GB/s \\
SRAM width & 512 bit & 512 bit & 512 bit & 512 bit \\
SRAM depth & 256 & 256 & 256 & 256 \\
SRAM banks & 256 & 256 & 1024 & 1024 \\
BF16 multipliers & 2112 & 8320 & 131584 & 131584 \\
BF16 adders & 2112 & 8320 & 131584 & 131584 \\
Miscellaneous PEs & 64 & 128 & 512 & 512 \\
Peak BF16 throughput\tnote{1}
& 4.224 TFLOP/s & 16.640 TFLOP/s
& 263.168 TFLOP/s & 263.168 TFLOP/s \\
\bottomrule
\end{tabularx}

\begin{tablenotes}[flushleft]\footnotesize
\item[] All configurations operate at 1 GHz. Device-matched
configurations use the largest valid PE organization whose peak
throughput does not exceed that of the target device. 
\textsuperscript{1} One multiplication and one addition count as two floating-point operations. Each PE issues one operation per cycle.
\end{tablenotes}
\end{threeparttable}

\section{LLaMA3-8B Results}
\label{app:llama70b}

\centering

\begin{threeparttable}
\begin{tabular}{@{}crrrrrrrr@{}}
\toprule
\multirow{2}{*}{\makecell{\textbf{Sequence}\\\textbf{length}}}
& \multicolumn{4}{c}{\textbf{Edge (batch 1)}}
& \multicolumn{4}{c}{\textbf{Server (batch 8)}} \\
\cmidrule(lr){2-5}\cmidrule(lr){6-9}
& \textbf{Prefill-TP} & \textbf{Prefill-EE}
& \textbf{Decode-TP} & \textbf{Decode-EE}
& \textbf{Prefill-TP} & \textbf{Prefill-EE}
& \textbf{Decode-TP} & \textbf{Decode-EE} \\
\midrule
128
& 4,002.75 & 317.05 & 126.27 & 7.44
& 238,287.84 & 669.32 & 7,560.52 & 53.41 \\
256
& 3,980.38 & 374.15 & 125.59 & 7.38
& 235,341.01 & 659.07 & 7,060.47 & 49.73 \\
512
& 3,952.80 & 394.61 & 124.29 & 7.28
& 226,956.91 & 638.69 & 6,243.70 & 43.78 \\
1024
& 3,908.76 & 384.40 & 121.84 & 7.09
& 200,261.22 & 606.23 & 5,097.66 & 35.52 \\
4096
& 3,667.25 & 306.37 & 108.75 & 6.16
& 130,890.76 & 459.65 & 2,586.75 & 17.78 \\
\bottomrule
\end{tabular}

\begin{tablenotes}[flushleft]\footnotesize
\item[] TP denotes throughput in GFLOP/s; EE denotes energy
efficiency in GFLOP/J.
\end{tablenotes}
\end{threeparttable}

\end{minipage}
\end{table*}

\clearpage

%% file: section/ae.tex
\section{Artifact Appendix}
\label{app:artifact}

\subsection{Abstract}

The artifact provides Accelgen, the compiler framework for \name{}. Starting
from PyTorch model code, the workflow generates Linalg-based MLIR, optimizes
Linalg operations, schedules kernels, and lowers the scheduled computation to
a Spatial Architecture Primitive (SAP) graph for performance modeling. The
outputs include intermediate MLIR and performance metrics such as latency,
energy, FLOPs, throughput, and energy efficiency. These outputs can be
inspected independently of the figures and tables in the paper.

\subsection{Artifact Checklist}

The artifact produces the following outputs to support reproducibility and
inspection of each compilation stage.
\begin{itemize}[leftmargin=2em,itemsep=0.25em]
  \item \textbf{Generated Linalg MLIR.} The Linalg representation generated
    from the input model.
  \item \textbf{Optimized Linalg MLIR.} The Linalg representation after
    Linalg-level optimizations.
  \item \textbf{Scheduled Clusters.} Computation clusters with their schedules
    and associated dataflow parameters.
  \item \textbf{SAP Graph.} The generated SAP graph representing the
    accelerator architecture.
  \item \textbf{Performance Report.} Modeled performance metrics for the
    generated accelerator.
\end{itemize}

\subsection{Availability and Installation}


The artifact is permanently archived on Zenodo and is available at
\url{https://doi.org/10.5281/zenodo.23085661}. The artifact can be
downloaded directly from Zenodo. A corresponding GitHub release is also
provided and matches the archived version:

\begin{verbatim}
$ git clone --branch ae \
https://github.com/XiaoyeAsZ/accelgen.git
\end{verbatim}

We provide two Docker-based methods for installing and running \name{}.

\textbf{Using the Prebuilt Docker Image (Recommended).} Pull the prebuilt
image and launch a container with the repository mounted as the working
directory.

\begin{verbatim}
$ cd accelgen
$ docker pull xiaoyeasz/accelgen:latest
$ docker run --gpus all -it --rm \
    -v $(pwd):/home/accelgen \
    -w /home/accelgen \
    xiaoyeasz/accelgen:latest
\end{verbatim}

\textbf{Building from the Dockerfile.} Alternatively, build the image locally
and launch a container using the same repository mount.

\begin{verbatim}
$ cd accelgen
$ docker build -f accelgen.dockerfile -t accelgen:latest .
$ docker run --gpus all -it --rm \
    -v $(pwd):/workspace/accelgen \
    -w /workspace/accelgen \
    accelgen:latest
\end{verbatim}

\subsection{Experimental Workflow}

The workflow consists of five stages that can be run separately to inspect
and reuse intermediate outputs. Each stage requires the output of the
preceding stage. Two evaluation modes are available: \texttt{rapid} for a
quick evaluation of the block~0 FFN workload in Llama3-8B, and \texttt{full}
for the complete evaluation suite. The full mode can take substantially
longer, particularly during kernel scheduling.

Inside the container (accelgen folder), build the compiler and enter the artifact evaluation directory:
\begin{verbatim}
$ cmake -S . -B build -G Ninja
$ cmake --build build
$ cd ae
\end{verbatim}

In the commands below, replace \texttt{MODE} with \texttt{rapid} or
\texttt{full}. All output directories are relative to
\texttt{ae/results/MODE/} in the repository.

\begin{enumerate}[leftmargin=2.4em,itemsep=0.45em]
  \item \textbf{Generate MLIR.}
  \begin{verbatim}
    $ bash MODE/01_generate_mlir.sh
  \end{verbatim}
  The script invokes \texttt{benchmark/dump\_mlir.py} for each selected
  combination of model, action, block, layer, batch size, and sequence
  length. The generated MLIR files are written to \texttt{mlir/}.

  \item \textbf{Optimize Linalg.}
  \begin{verbatim}
    $ bash MODE/02_linalg_opt.sh
  \end{verbatim}
  The optimizer performs mixed-precision resolution, named-operation
  generalization, generic marking, dead-operation elimination, generic
  fusion, and tensor-operation folding. The optimized MLIR files are
  written to \texttt{generic/}.

  \item \textbf{Schedule Kernels.}
  \begin{verbatim}
    $ bash MODE/03_kernel_schedule.sh
  \end{verbatim}
  The \texttt{func.func(kernel-schedule\{...\})} pass searches over tiling,
  loop, unrolling, and cluster choices using the selected edge or server
  resource configuration. The scheduled MLIR files are written to
  \texttt{scheduled/}. By default, each cluster contains at most six
  operations (\texttt{MAX\_OPERATION=6}).

  \item \textbf{Construct SAP graph.}
  \begin{verbatim}
    $ bash MODE/04_lower_to_dap.sh
  \end{verbatim}
  The \texttt{convert-to-dap\{config-path=...\}} pass lowers each scheduled
  workload to the hardware-facing DAP dialect. The resulting MLIR files
  are written to \texttt{dap/} and contain operations such as
  \texttt{dap.sram}, \texttt{dap.data\_path}, arithmetic operations, and
  routing operations.

  \item \textbf{Model Performance.}
  \begin{verbatim}
    $ bash MODE/05_model_performance.sh
  \end{verbatim}
  The \texttt{model-performance\{config-path=...\}} pass estimates latency,
  energy, and FLOPs and emits annotated MLIR files in \texttt{performance/}.
  The accompanying \texttt{logs/performance.log} contains workload headers
  and metric records suitable for processing with the repository's
  analysis scripts.
\end{enumerate}